\documentclass[useAMS,usenatbib]{mn2e}
\usepackage{graphicx} 
\usepackage{amsmath} 
\usepackage{todonotes}

\title[Bayesian power spectrum estimation]{Bayesian power spectrum estimation at the Epoch of Reionization}
\author[P. Sims et al.]{\parbox{\textwidth}{P. H. Sims$^{1}$\thanks{E-mail:
peter_sims1@brown.edu}, L. Lentati$^{2}$, J. C. Pober$^{1}$, C. Carilli$^{3,2}$, M. P. Hobson$^{2}$, P. Alexander$^{2}$, P. Sutter$^{4,5,6}$}\vspace{0.4cm}\\ %
$^{1}$Department of Physics, Brown University, Providence, RI 02912, USA \\
$^{2}$Astrophysics Group, Cavendish Laboratory, JJ Thomson Avenue,  Cambridge, CB3 0HE, UK\\
$^{3}$National Radio Astronomy Observatory, Socorro, NM 87801, USA\\
$^{4}$INFN - National Institute for Nuclear Physics, via Valerio 2, I-34127 Trieste, Italy\\
$^{5}$INAF - Osservatorio Astronomico di Trieste, via Tiepolo 11, 1-34143 Trieste, Italy\\
$^{6}$Center for Cosmology and Astro-Particle Physics, Ohio State University, Columbus, OH 43210}

\begin{document}

\maketitle
\label{firstpage}

\begin{abstract}
We detail a new method for performing robust Bayesian estimation of the three-dimensional spatial power spectrum of the Epoch of Reionization (EoR) from interferometric observations.  The versatility of this technique allows us to present two approaches.  First, when the observations span only a small number of independent spatial frequencies ($k$-modes) we sample directly from the spherical power spectrum coefficients that describe the EoR signal realisation.  Second,  when the number of $k$-modes to be included in the model becomes large, we sample from the joint probability density of the spherical power spectrum and the signal coefficients, using Hamiltonian Monte Carlo methods to explore this high dimensional ($\sim$ 20000) space efficiently.  This approach has been successfully applied to simulated observations that include astrophysically realistic foregrounds in a companion publication (\citealt{2016MNRAS.462.3069S}). Here we focus on explaining the methodology in detail and use simple foreground models to both demonstrate its efficacy and highlight salient features.  In particular, we show that including an arbitrary flat spectrum continuum foreground that is $10^8$ times greater in power than the EoR signal has no detectable impact on our parameter estimates of the EoR power spectrum recovered from the data.
\newline
\newline
\end{abstract}


\section{Introduction}
  
The Epoch of Reionization (EoR) marks a period of history that began approximately 400 Myr after the Big Bang, when the first ionizing sources formed in an otherwise neutral Universe.  For a detailed review of the EoR refer to, for example, \cite{2012RPPh...75h6901P}, \cite{2013fgu..book.....L} and \cite{2010ARA&A..48..127M}. In brief, the emergence of these sources resulted in the gradual formation of ionized `bubbles' in the neutral hydrogen that made up the surrounding intergalactic medium. These bubbles are thought to have expanded over a redshift range from $z \sim 16$ to 6; however, the precise timing and duration of the period, as well as the spatial scales on which these bubbles evolved, are questions that largely remain unanswered.

Recent observations have been able to constrain the bright end of the galaxy luminosity function at low redshifts ($ z \la 8$; \citealt{2010ApJ...709L.133B,2013ApJ...768..196S}), and other observational programmes have placed constraints on reionization, for example, from the optical depth of Thompson scattering to the CMB \citep{2014A&A...571A..16P}. However, the most promising probe for answering these questions more completely may lie in the detection of the redshifted 21-cm signal from the EoR, since it provides a direct link to the density and distribution of the neutral hydrogen during that time.  

The wealth of information encoded in the 21-cm signal has meant that its detection is one of the major goals of existing and upcoming low frequency interferometers, such as the Giant Metrewave Radio Telescope (GMRT; \citealt{2013MNRAS.433..639P}),  the LOw Frequency ARray (LOFAR; \citealt{2013A&A...556A...2V}), the Murchison Widefield Array (MWA; \citealt{2013PASA...30....7T}), the Precision Array to Probe the Epoch of Reionization (PAPER; \citealt{2014ApJ...788..106P}), the Hydrogen Epoch of Reionization Array (HERA; \citealt{2014ApJ...782...66P,2017PASP..129d5001D}) and the Square Kilometre Array (SKA; \citealt{2013ExA....36..235M}).
Our principal focus in this paper will be the extraction of information from interferometric observations of the `late stage' ($z \sim 6\to10$) 21-cm EoR power spectrum, averaged over either 2 or 3 dimensions of Fourier space to form `cylindrical' or `spherical' power spectra, respectively.  

In recent years, Bayesian methods have become more prevalent in the analysis of interferometric data sets, both in terms of providing optimal imaging techniques \citep{2014MNRAS.438..768S} and power spectrum analysis in the context of the cosmic microwave background (e.g. \citealt{2012ApJS..202....9S, 2013ApJS..204...10K}).
In the case of the EoR, in order to determine the power spectrum of the fluctuations robustly, one must also account for the presence of Galactic and extragalactic foreground emission (see e.g. \citealt{1999A&A...345..380S, 2005ApJ...625..575S,2013ApJ...768L..36P}), which can be orders of magnitude greater than the EoR signal of interest.  
The last decade has seen a significant number of techniques developed to remove, or otherwise mitigate, these foregrounds before the estimation of the cosmological signal (e.g. \citealt{2009ApJ...695..183B,2009MNRAS.397.1138H,2012MNRAS.423.2518C,2013MNRAS.429..165C,2015MNRAS.447.1973B,2017arXiv171110834M}).  Ideally, however, one would fit simultaneously for the EoR signal and the foregrounds in order to account for the covariance between the two and, hence, produce an unbiased estimate of the EoR power spectrum.

In this paper we outline a general Bayesian framework that allows just such a joint analysis and we consider two different regimes.
First, when the number of spatial scales we wish to include in our model for the EoR power spectrum is small ($\la 20000$), we sample directly from the spherical power spectrum coefficients of the EoR signal. This results in a computational problem that is low--dimensional ($\sim 10$), but high in computational expense, with large, dense matrix inversions required in every likelihood calculation. As such, when the number of spatial scales to be included is larger and the matrix inversions required by the analysis become computationally intractable, we sample instead from the joint probability density of the spherical power spectrum coefficients and the EoR signal realisation.  This allows us to eliminate all matrix-matrix multiplications and costly matrix inversions from the likelihood calculation entirely, replacing them with matrix-vector operations and diagonal matrix inversions. In this case, the dimensionality is much larger ($\sim 20000$) and so we perform the sampling process using a Guided Hamiltonian Sampler (GHS; Balan, Ashdown $\&$ Hobson, in prep, henceforth B18; see also e.g. \citealt{2013PhRvD..87j4021L} for uses in other astrophysical fields), which exploits Hamiltonian Monte Carlo sampling methods to provide an efficient means of sampling in large numbers of dimensions (potentially $\sim 10^6$).  This framework has been applied to simulated interferometric observations that combined realistic astrophysical foregrounds and the EoR signal in \cite{2016MNRAS.462.3069S} where it was shown to result in unbiased estimates of the three-dimensional power spectrum of the EoR for $\log_{10}(k[h\mathrm{Mpc^{-1}}]) > -1.0$. In this work, we emphasize a detailed explanation of this methodology, using simple foreground models to both demonstrate its efficacy and to highlight salient features.

The remainder of this paper is organized as follows.
In Sections \ref{Section:Interferometry} - \ref{Section:Marginalisation}, we derive the likelihood functions for both the cases of small and large dimensionality.  In Section \ref{Section:Sampling}, we describe the guided Hamiltonian sampler and how it can be applied to interferometric data analysis.  In Section \ref{Section:Sims}, we then apply this framework to a set of simulations to demonstrate the efficacy of the method.  These include both high and low signal-to-noise data sets, with and without an additional flat spectrum continuum component.  In the latter case, we show that this extra continuum component does not affect the power spectrum estimation of the EoR signal present in the data set, despite assuming no prior knowledge of the distribution or amplitudes of sources. 
In Section \ref{Section:Comparison}, we compare the methods presented here with other 21-cm power spectrum estimators in the literature, including the Bayesian approaches of \cite{2015MNRAS.452.1587G} and \cite{2016ApJS..222....3Z}.
Finally, in Section \ref{Section:Conclusions}, we offer some concluding remarks.

\section{Observing with an interferometer}
\label{Section:Interferometry}

For a generic radio interferometer, the Measurement Equation \citep{1996A&AS..117..137H, 2011A&A...527A.106S}, for a pair of antennas $p,q$ observing a single point source allows us to construct a `visibility matrix', $\mathbfss{V}_{pq}$, as,

\begin{equation}
\label{Eq:MEq}
\mathbfss{V}_{pq} = \mathbfss{J}_{p}\mathbfss{B}\mathbfss{J}_{q}^\mathbf{H},
\end{equation}
where $\mathbf{H}$ denotes the Hermitian transpose,  $\mathbfss{B}$ is the `brightness matrix,' given by,

\begin{equation}
\mathbfss{B} = \begin{bmatrix}
I + Q   & U + iV \\ 
U - iV  & I - Q
\end{bmatrix},
\end{equation}
for Stokes parameters $(I, Q, U, V)$, and $\mathbfss{J}_{p}$ and $\mathbfss{J}_{q}$ are $2\times 2$ Jones matrices that describe the cumulative product of  all propagation effects along the signal path.

In this work we will be considering only observations that are uncorrupted by (or have been corrected for), for example, calibration errors or ionospheric effects. Thus, the only contributions to the Jones matrices that we will consider are those that come from a scalar phase delay $K_p$ for each antenna $p$, defined,

\begin{equation}
K_p = \exp\left(-2\pi i(u_pl + v_pm + w_p(n-1)\right),
\end{equation}
with $l,m$ and $n = \sqrt{1-l^2-m^2}$ the direction cosines of the unit vector, $\mathbf{\hat{r}}$, from the antenna to the sources and $(u,v,w)$ the antenna coordinates in wavelengths. Integrating over the whole sky, we can therefore rewrite Eq. \ref{Eq:MEq}, explicitly including only the discussed terms, as,
\begin{eqnarray}
\label{Eq:MEq2}
\mathbfss{V}_{pq} &=& \int_0^{4\pi} \mathrm{d}\Omega \mathbfss{P}_{p}(\mathbfit{x})\mathbfss{B}(\mathbfit{x})\mathbfss{P}_{q}^\mathbf{H}(\mathbfit{x})\nonumber \\
&\times& \exp\left(-2\pi i(\mathbfit{u}_{pq}\cdot\mathbfit{x})\right),
\end{eqnarray}
with $\mathbfss{P}_{p}$ a term that describes the voltage beam pattern of antenna $p$, $\mathbfit{x} = (l,m,\sqrt{1-l^2-m^2})$, and $\mathbfit{u}_{pq} = (u_p - u_q,v_p - v_q,w_p - w_q)$.

In practice, this integral is difficult to evaluate directly, and so we perform a sine projection onto the plane $(l,m)$ at the field centre.  If the field of interest is sufficiently small, we can also make the approximation that,
\begin{equation}
\left(\sqrt{1-l^2-m^2} - 1\right)w \approx -\frac{1}{2}(l^2+m^2)w \approx 0
\end{equation}
and so consider only $\mathbfit{l} = (l,m)$, $\mathbfit{u}_{pq} = (u,v)$.  
In the context of the simulations presented in Section \ref{Section:Sims}, we consider a primary beam with a  full width at half maximum of 8 degrees.  For an 8 degree separation, we would have $l^2+m^2 \sim 0.02$, which we consider to be in the regime where this approximation holds,  resulting in the expression,

\begin{eqnarray}
\label{Eq:MEq3}
\mathbfss{V}_{pq} &=& \int\mathrm{d}^2\mathbfit{l} \; \mathbfss{P}_{p}(\mathbfit{l} )\mathbfss{B}(\mathbfit{l} )\mathbfss{P}_{q}^\mathbf{H}(\mathbfit{l})\nonumber \\
&\times& \exp\left(-2\pi i(\mathbfit{u}_{pq}\cdot\mathbfit{l})\right).
\end{eqnarray}

Finally if we consider only the total intensity of the sky $I(\mathbfit{l}, \nu)$, we obtain, for any pair of antennas (or `baseline', $i$) operating at a single frequency $\nu$, the expression,

\begin{equation}
\label{Eq:viseq}
V_i(\mathbf{u}_i,\nu) = \int\mathrm{d}^2\mathbfit{l} \;P_i(\mathbfit{l},\nu)I(\mathbfit{l},\nu)\exp(-2\pi i\mathbf{u}_i\cdot\mathbfit{l}),
\end{equation}
where we have dropped the subscript $\mathbfit{pq}$ for the coordinate vector $\mathbf{u}_i$, and we have replaced the visibility matrix $\mathbfss{V}_{pq}$ with the complex number $V_i(\mathbf{u}_i,\nu)$ and the product $\mathbfss{P}_{p}\mathbfss{P}_{q}$ with $P_i(\mathbfit{l},\nu)$, the primary beam profile for baseline $i$.  We have also written all quantities explicitly as a function of the observing frequency $\nu$.

By the convolution theorem, which states that the Fourier transform of a product of functions is the convolution of the Fourier transforms of the functions separately, we can define the aperture function $A(\mathbf{u}, \nu)$ as the Fourier transform of the primary beam $P(\mathbfit{l},\nu)$ and the complex visibility plane $S(\mathbf{u},\nu)$ as the Fourier transform of the sky brightness $I(\mathbfit{l},\nu)$.  Therefore Eq. \ref{Eq:viseq} can be rewritten as,

\begin{equation}
\label{Eq:convvis}
V_i(\mathbf{u}_i,\nu) = \int\mathrm{d}^2\mathbf{u} \;A(\mathbf{u}-\mathbf{u}_i,\nu) S(\mathbf{u},\nu).
\end{equation}

In the following sections, we now describe our model for $S(\mathbf{u},\nu)$ that allows us to reconstruct the observed visibilities $V_i(\mathbf{u_i},\nu)$, while remaining computationally tractable to evaluate.

\subsection{Constructing a likelihood}

We begin by considering the complex visibilities obtained during an interferometric observation to be the sum of a signal component $\mathbf{s}$ sampled from the visibility plane and an instrumental noise component $\mathbf{n}$, where we describe the noise as a zero--mean statistically homogeneous Gaussian random field, uncorrelated between different visibilities, with covariance matrix $\mathbfss{N}$  given by,

\begin{equation}
N_{ij} = \left< n_in_j^*\right> = \delta_{ij}\sigma_{j}^{2} \ ,
\end{equation}
where $\left< .. \right>$ represents the expectation value and $\sigma_{j}$ is the rms value of the noise term for visibility $j$. 

We can, therefore, write our data vector $\mathbf{d}$ containing $N_{\mathrm{vis}}$ complex visibilities as,

\begin{equation}
\mathbf{d} = \mathbf{s} + \mathbf{n},
\end{equation}
allowing us to construct a general likelihood for a model vector $\mathbf{m}$ constructed from the set of parameters $\mathbf{\Theta}$ as,

\begin{eqnarray}
\label{Eq:BasicVisLike}
\mathrm{Pr}(\mathbf{d} | \mathbf{\Theta}) &=& \frac{1}{\sqrt{(2\pi)^{2N_{\mathrm{vis}}})\mathrm{det}\mathbfss{N}}} \nonumber \\
&\times&\exp\left[-\frac{1}{2}\left(\mathbf{d} - \mathbf{m(\Theta)}\right)^T\mathbfss{N}^{-1}\left(\mathbf{d} - \mathbf{m(\Theta)}\right)\right].
\end{eqnarray}
Henceforth, for clarity in the mathematical notation, we will consider our data and, hence, model vectors to be the concatenation of the real part and imaginary parts, rather than a set of complex values.  As such $\mathbf{d}$ and $\mathbf{m}$ will be vectors of length $2\times N_{\mathrm{vis}}$, while the diagonal elements of $\mathbfss{N}$ will be given by the variance of the Gaussian noise in the real and imaginary parts of the observed visibilities separately. 

\subsection{A model grid}
\label{Section:Grid}

In the context of Eq. \ref{Eq:convvis}, our model $\mathbf{m(\Theta)}$ will be a representation of the complex visibility plane $S(\mathbf{u},\nu)$.  In principle, we can simply divide the complex plane into equal area cells of side $\Delta u$. However, our ability to  determine the properties of the power spectrum correctly will clearly depend on the number of cells chosen to make up our model.  Since both the number of samples required and the speed of the likelihood evaluation will be strongly dependent upon the number of cells used, a compromise must be made between how accurately our model can represent the true complex plane and our ability to perform the computational analysis.  

In practice, a natural maximum size for the model cells exists, as an antenna of diameter $D$ will convolve the complex visibility plane with a function that has scale length $\sim D/2\lambda$.  As such, from sampling theory, we require $\Delta u < D/2\lambda$ in order for our model to adequately describe the underlying visibility plane.

As we are working with a model grid, initially it may seem  a more natural choice to define our model in the image domain.  Here, we can construct a uniformly spaced $N_{\mathrm{pix}}\times N_{\mathrm{pix}}\times N_{\mathrm{chan}}$ cube, where $N_{\mathrm{pix}}$ is the number of pixels along one side of the image and $N_{\mathrm{chan}}$ is the number of frequency channels.  Defining the vector of model amplitudes for the pixels in the image as $\mathbf{c}$, we can then generate a set of model visibilities $\mathbf{m}$ as,

\begin{equation}
\mathbf{m} = \mathbfss{F}_\mathbf{n}^{-1}\mathbfss{P}\mathbf{c},
\end{equation}
where $\mathbfss{P}$ is a diagonal $N_{\mathrm{pix}}\times N_{\mathrm{pix}} \times N_{\mathrm{chan}}$ matrix that encodes the primary beam correction and $\mathbfss{F}_\mathbf{n}^{-1}$ is a $2N_{\mathrm{vis}}\times N_{\mathrm{pix}}^2N_{\mathrm{chan}}$ matrix, where the factor 2, as previously discussed,  accounts for the real and imaginary parts and describes the inverse Fourier transform from our primary beam corrected image, to the sampled $(u, v)$ coordinates.

%

%

%

As we will show in Section \ref{Section:IntLike}, however, when we include a prior on the EoR signal, that prior is defined in $k$-space, such that contributions to the signal from terms with similar $|k|$ values will be considered to come from a single Gaussian distribution of some variance to be determined during the analysis.  When defining the prior in this way, constructing our model in the image domain results in large, dense matrix inversions, rapidly making the analysis intractable.  If we construct our model in the UV domain instead, this prior matrix becomes diagonal and the inversions become trivial to compute.

In principle, we could replicate the effect of the primary beam in the UV domain by performing a convolution of our model UV grid with the telescope aperture function, the Fourier transform of the primary beam.  This, however, is much less computationally tractable than performing a simple multiplication in the image domain and then performing a Fourier transform of this primary-beam-corrected image to the UV domain. We can  combine the speed of defining our model in the UV domain with the efficiency of performing the primary beam correction in the image domain, by defining a new matrix $\bar{\mathbfss{F}}$, which acts on a vector of model parameters $\mathbf{a}$, where $\mathbf{a}$ describes the amplitudes for the real and imaginary parts for a grid of points in the UV-plane, such that our model visibilities $\mathbf{m}$ are given by,

\begin{eqnarray}
\label{Eq:finalP}
\mathbf{m} &=& \bar{\mathbfss{F}}\mathbf{a} \\ \nonumber
&=& \mathbfss{F}_\mathbf{n}^{-1}\mathbfss{P}\mathbf{F}\mathbf{a} \ ,
\end{eqnarray}
where the matrix $\mathbf{F}$ is simply the Fourier transform from our grid of $(u,v)$ domain points, to the grid of image domain pixels.  This additional multiplication has no impact on the evaluation time of our likelihood, as we can simply precompute the matrix product $\mathbfss{F}_\mathbf{n}^{-1}\mathbfss{P}\mathbf{F}$ and still evaluate the model vector $\mathbf{m}$ in a single matrix--vector multiplication.

\subsection{Including large spatial scales}
\label{Section:largeScales}

While our definition of the matrix $\mathbf{F}$ in Eq. \ref{Eq:finalP} represents a standard 2--dimensional Fourier transform, this will not correctly model power on spatial scales greater than the size of the image, or equivalently, on scales with $|\mathbf{u}| < D/2\lambda$.  Linear trends that extend across the image, such as those that can be expected from Galactic foregrounds, will `leak' into the model coefficients that describe power on scales less than the image size, with $|\mathbf{u}| > D/2\lambda$.

In principle, we could incorporate these scales into our model by simply reducing the cell size in our UV model, $\Delta u$.  For example, power on scales 10 times the size of the image could be incorporated robustly simply by choosing a cell size of $0.1 \times \Delta u$.  This, however, is not computationally tractable, as it will increase the dimensionality of our problem by a factor of 100. In the context of one--dimensional power spectrum recovery, it has been shown by van Haasteren et al. (2014) that, for a data vector of length $T$, using a log spacing for sub-harmonic frequencies ($\nu < 1/T$), and linear spacing for $\nu \geq 1/T$ in steps of $\Delta \nu = 1/T$, allows for accurate recovery of the power spectrum, when there is significant power in these low frequency terms.

We, therefore, take an equivalent approach with our two--dimensional analysis.  We define a set of 10 evenly log spaced spatial scales between the size of the image, and 10 times the size of the image, and include them in our model, simultaneously with the linear UV domain grid with cell size $\Delta u$.

The matrix $\mathbf{F}$, therefore, no longer represents the transform from a uniform grid of UV domain model points to the uniform grid of image pixels.  Instead, it defines the transform from our complete  UV model, including the points describing power at large spatial scales, to the uniform grid of image pixels.  With this redefinition of $\mathbf{F}$, Eq.  \ref{Eq:finalP} remains unchanged.

\subsection{Incomplete UV coverage}

In any interferometric observation, the coverage of the UV-plane will not be complete.  In particular, an interferometer is not sensitive to the (0,0) UV coordinate, as the minimum separation between two antennas cannot be less than the size of the dish, $D$, so that any observation made will be insensitive to the true mean of the sky.  More generally, however, the sampling of the UV plane by our interferometer will result in gaps, or areas of decreased sensitivity, the precise nature of which will be determined by the arrangement of antennas and length of observation.

We can compute the weighting in the UV--plane that results from the sampling of a set of $N_{\mathrm{vis}}$ discrete visibilities, which can also be considered the Fourier transform of the interferometer point spread function. For the current analysis, we compute these weights by defining a gridding matrix $\mathbfss{G}$ as,

\begin{equation}
\mathbfss{G} = \mathbfss{F}^{-1}\mathbfss{F}_\mathbf{n},
\end{equation} 
where $\mathbfss{F}_\mathbf{n}$ is a $N_{\mathrm{pix}}^2\times 2N_{\mathrm{vis}}$ matrix representing the direct Fourier transform of the visibilities to an $N_{\mathrm{pix}}\times N_{\mathrm{pix}}$ image domain grid, and $\mathbfss{F}^{-1}$ describes the Fourier transform from the image grid to our grid of UV cells.

%

In principle, UV cells far from the points sampled by the interferometer could have non--zero weights, but would contribute a negligible amount to our model. It is therefore of interest to calculate the weight of any UV cell, $W_j$. The covariance matrix of the weighted visibilities projected onto the space of our gridded $uv$-model is given by,
\begin{equation}
\label{Eq:Weight}
\mathbfss{W} = \mathbfss{G}\mathbfss{N}^{-1}\mathbfss{G}^T.
\end{equation}
In this work, we approximate the weight of UV cells $j$ by the $j$-th diagonal of the weight matrix, $W_{jj}$, and consider the $N_{\mathrm{uv}}$ element subset of highest-weighted cells summing to 99$\%$ of the total weight in the definition of our matrix $\mathbfss{F}$.

Working directly in the UV domain, it is therefore straightforward to account for the incomplete UV coverage of a given observation.

\subsection{The full $k$-cube}
\label{Section:KCube}


As mentioned in Section \ref{Section:Grid} we define our EoR signal model directly in $(k_x, k_y, k_z)$ space, using the set of $(k_x, k_y)$ points that correspond to the set of $N_{\mathrm{uv}}$ gridded UV coordinates that we include in our analysis. These $(u,v)$ coordinates can be translated directly to $k_x, k_y$ coordinates through the relations,

\begin{eqnarray}
k_x &=& \frac{2\pi u}{D_m}  \\ \nonumber
k_y &=& \frac{2\pi v}{D_m}   \\ \nonumber \ ,
\end{eqnarray} 
with $D_m$ the transverse comoving distance from the observatory to the redshift of the EoR observation. 

Our sampled visibilities, however, are defined in $(\bmath{u},\nu)$ space therefore we must also transform our model cube from $k_z$ to observing frequency $\nu$. We first define the matrix $\mathbfss{F}_z$ as,
\begin{equation}
\label{Eq:FMatrix}
F_z(\nu, n_{\eta}) = \frac{2\nu^2\kappa_B}{10^{-23}c^2}\frac{\sqrt{2}}{N_{\mathrm{chan}}}\sin\left(\frac{2\pi}{B}n_{\eta} \nu\right),
\end{equation}
with an equivalent cosine term, $B$ the bandwidth of the observation and $\eta = n_{\eta}/B$ the Fourier domain parameter after transforming along the frequency axis, where we include terms up to some maximum $n_{\eta} = n_{\mathrm{max}\eta}$, with $n_{\mathrm{max}\eta}$ determined by the bandwidth of the dataset and constrained such that the number of data points minus the total model parameters is non-negative. The factor $2\nu^2\kappa_B/c^2$ from the Rayleigh-Jeans law, with $\kappa_B$ the Boltzmann constant and $c$ the speed of light,  at the front of this expression allows us to convert from units of mK  in the model $(k_x, k_y, k_z)$ cube, to Janskys (1 Jy = $10^{-26}$ Wm$^{-2}$Hz$^{-1}$). We can then relate $\eta$ to the cosmological parameter $k_z$ via the relation,
\begin{equation}
k_z = \frac{2\pi H\mathrm{_0}f_{21}E(z)}{c(1+z)^2}\eta,
\end{equation}
with $z$ the redshift of the EoR observation, $H\mathrm{_0}$ the Hubble constant, $E(z)$ the dimensionless Hubble parameter, $f_{21}$ the frequency of the 21cm line emission and $c$ the speed of light.

While $\mathbfss{F}_z$ represents a typical 1D Fourier transform, the EoR signal present in the data will include fluctuations on scales much longer than the bandwidth of the observation.  Written as in Eq. \ref{Eq:FMatrix}, this transform will not correctly account for these low frequencies, causing them to `leak' into the higher frequency terms included in our model, biasing the power spectrum parameter estimates at the scales of interest.  In principle, these low frequencies could be included simply by adding additional log--spaced Fourier modes with $n_{\eta} < 1$ to $\mathbfss{F}_z$; as in Section \ref{Section:largeScales}, using a log spacing for the sub-harmonics allows for accurate recovery of the spectrum when there is significant power in these low frequency terms. In  van Haasteren et al. (2014) they note, however, that except for the most extreme cases, these sub-harmonics terms can also be well modelled by a simple quadratic in frequency.  

In our model, we therefore  include a quadratic in frequency to act as a proxy to the subharmonic structure in our data, 
\begin{equation}
\label{Eq:Quad}
Q_z = \frac{2\nu_o^2\kappa_B}{10^{-23}c^2}(q_0 + \nu q_1 + \nu^2q_2) \equiv \mathbfss{Q}_z\bmath{q},
\end{equation}
where $\bmath{q} = (q_0, q_1, q_2)$ are amplitude parameters to be fit for.

We can therefore write our final model, given by the concatenated $N_{\mathrm{uv}}\times n_{\mathrm{max}\eta}$ length vector of signal coefficients $\bmath{a}$ and $N_{\mathrm{uv}}\times 3$ quadratic coefficients $\bmath{q}$ defined in the k-cube, as,
\begin{equation}
\label{Eq:FinalModel}
\mathbf{m} = \bar{\mathbfss{F}}\left(\mathbfss{F}_z\bmath{a} + \mathbfss{Q}_z\bmath{q}\right) \ .
\end{equation}
Here, $\mathbfss{F}_z$ and  $\mathbfss{Q}_z$ both now represent block diagonal matrices that act independently on each set of coefficients $(\mathbf{a_i}, \mathbf{q_i})$, for each model UV cell $i$, and $\bar{\mathbfss{F}}$ is the two--dimensional primary beam corrected transform described in Eq. \ref{Eq:finalP}.  Our likelihood at this stage becomes, 

\begin{eqnarray}
\label{Eq:DataLike}
\mathrm{Pr}(\mathbf{d} | \bmath{a},\bmath{q}) &=& \frac{1}{\sqrt{(2\pi)^{2N_{\mathrm{vis}}})\mathrm{det}\mathbfss{N}}} \nonumber \\
&\times&\exp\left[-\frac{1}{2}\left(\mathbf{d} - \bar{\mathbfss{F}}\left(\mathbfss{F}_z\bmath{a} + \mathbfss{Q}_z\bmath{q}\right)\right)^T \right.\nonumber\\ 
&\times&\left.\mathbfss{N}^{-1}\left(\mathbf{d} - \bar{\mathbfss{F}}\left(\mathbfss{F}_z\bmath{a} + \mathbfss{Q}_z\bmath{q}\right)\right)\right].
\end{eqnarray}

\subsection{Including foreground models}
\label{Section:Foregrounds}

In order to make a detection of the EoR power spectrum, correctly accounting for foreground signals in the visibilities will be key. These include diffuse emission from the Galaxy and continuum emission from extragalactic sources (e.g. \citealt{2008MNRAS.389.1319J}), which, in combination, can be up to five orders of magnitude greater than the EoR signal of interest \citep{1999A&A...345..380S}. 

In principle, any additional foreground model, $\mathbf{m}(\Theta_{\mathrm{fg}})$, can be added to the model in Eq \ref{Eq:DataLike}, either in the image domain, or in the UV. In this case we can write the data likelihood as,

\begin{eqnarray}
\label{Eq:DataLikeFG}
\mathrm{Pr}(\mathbf{d} | \mathbf{a}, \mathbf{\Theta_{fg}}) &=& \frac{1}{\sqrt{(2\pi)^{2N_v})\mathrm{det}\mathbfss{N}}} \\ \nonumber
&\times&\exp\left[-\frac{1}{2}\left(\mathbf{d} - \bar{\mathbfss{F}}(\mathbfss{F}_z\mathbf{a}+\mathbfss{Q}_z\bmath{q}) - \mathbf{m}(\Theta_{\mathrm{fg}})\right)^T \right.\nonumber \\
&\times&\left.\mathbfss{N}^{-1}\left(\mathbf{d} - \bar{\mathbfss{F}}(\mathbfss{F}_z\mathbf{a}+\mathbfss{Q}_z\bmath{q}) - \mathbf{m}(\Theta_{\mathrm{fg}})\right)\right]
\end{eqnarray}
and then proceed to sample over the joint parameter space ($\mathbf{a}, \mathbf{q}, \mathbf{\Theta_{\mathrm{fg}}}$).

One approach advocated to  model smooth foreground emission is to use a simple polynomial in frequency (e.g. \citealt{2009ApJ...695..183B}). In Section \ref{Section:KCube} we note that we include a quadratic in our Fourier transform from frequency to the parameter $\eta$ in order to model any low frequency variations that exist in the data with periods longer than the bandwidth of the observation. This quadratic term can therefore serve as a rudimentary model for the foregrounds in our analysis; we reiterate, however, that the primary purpose of the quadratic is simply to provide us with an unbiased estimate of the scales of interest (i.e. with $n_{\eta} \ge 1$ in Eq. \ref{Eq:FMatrix}).  In principle, higher order terms could also be added, however these will be increasingly covariant with the Fourier modes included in the model, and so we do not take this approach. In future work, we will explore the inclusion of astrophysically motivated foreground models to better separate foreground signatures from the EoR.

In Section \ref{Section:IntLike}, we describe our approach to estimating the EoR power spectrum, including a prior on the signal coefficients $\mathbf{a}$ that incorporates the assumption that the EoR signal is spatially homogenous. We do not, however, incorporate such a prior on the quadratic terms in our estimation of the power spectrum, as these terms will likely be dominated by foreground emission and, so, will not have the same homogeneity, at least in the case of the Galactic foreground. While, in principle, a separate Gaussian prior could be included for these quadratic terms, in order to make our analysis of the EoR signal more conservative, we use a less informative uniform prior on the amplitudes of these coefficients.

In our simulations in Section \ref{Section:Sims} we will be considering only simple continuum models, with flat spectrum sources; however, a more detailed account on the effect of foregrounds, when including realistic frequency evolution and spatial structure, using the technique described in this work is given in \cite{2016MNRAS.462.3069S}.

\section{Estimating the Power Spectrum}
\label{Section:IntLike}

Assuming the EoR signal to be spatially homogenous, the covariance matrix $\bmath{\Phi}$ of the $k$-space coefficients $\bmath{a}$ will be diagonal, with components,
\begin{equation}
\label{Eq:Prior}
\Phi_{ij} = \left< a_ia_j\right> = \varphi_{i}\delta_{ij} \ ,
\end{equation}
where there is no sum over $i$, and the set of coefficients $\varphi_i$ represent the theoretical power spectrum for the EoR signal.

In the framework we will describe below, we are free to choose any functional form for the coefficients $\varphi_i$.  It is here then that, should one wish to fit a specific model to the power spectrum at the point of sampling -- to perform model selection, for example -- the set of coefficients $\varphi_i$ should be given by some function $f(\Theta)$, where we sample from the parameters $\Theta$ from which the power spectrum coefficients $\varphi_i$ can then be derived. 

In Section \ref{Section:Sims} we will be comparing the results of our method with an input simulation obtained using the seminumerical 21cm FAST algorithm \citep{2011MNRAS.411..955M,2007ApJ...669..663M}.  After computing the EoR simulation, 21cmFAST outputs a spherical power spectrum of the simulated cube, performing a 3--dimensional FFT and averaging all the Fourier coefficients that fall within some spherical shell in $k$-space in order to calculate the power spectrum within that bin.  In order to draw the most direct comparison with the input simulation, we therefore calculate the quantity $|k| = \sqrt{k_x^2 + k_y^2 + k_z^2}$ for each $k$--space coefficient $\mathbf{a}$ in our model and define a set of bins in the quantity $|k|$.  As in 21cmFAST, we define the edges of these bins to be spaced as $1.5^n\Delta|k|$ for bins $n=1\dots n_{\mathrm{max}}$, with  $n_{\mathrm{max}}$ the largest bin included in the model.  Our model for the power spectrum $\mathbf{\varphi}$ will then be a set of independent parameters $\varphi_{i}$, one for each $|k|$ bin $i$.

We then write the joint probability density of the model coefficients that define our power spectrum and the $k$-space signal coefficients Pr$(\bmath{\varphi}, \mathbf{a} \;|\; \mathbf{d})$ as,

\begin{equation}
\label{Eq:Prob}
\mathrm{Pr}(\bmath{\varphi}, \mathbf{a} , \mathbf{q}\;|\; \mathbf{d}) \; \propto \; \mathrm{Pr}(\mathbf{d} | \mathbf{q}, \mathbf{a}) \; \mathrm{Pr}(\mathbf{a} | \bmath{\varphi}) \; \mathrm{Pr}(\bmath{\varphi}) \; \mathrm{Pr}(\bmath{q})
\end{equation}
and then marginalise over all $\mathbf{a}$ and  $\mathbf{q}$  in order to find the posterior for the parameters that define the power spectrum $\bmath{\varphi}$ alone.

For our choice of $\mathrm{Pr}(\bmath{\varphi})$, we use either a uniform prior in the amplitude of the coefficient, or a uniform prior in $\log_{10}$ space.  The latter case is the least informative prior we can choose; however, when the goal is to set an upper limit on 21cm emission, a prior that is uniform in log space is not appropriate, as the upper limit is dependent upon the bounds of the prior. When this is the case, we use a prior that is uniform in the amplitude. In either case, we draw our samples from the parameter $\rho_i$, such that,

\begin{equation}
\varphi_i = \frac{2\pi^2N_{\mathrm{pix}}^2N_{\mathrm{chan}}\Omega_{\mathrm{pix}}^4}{|\bmath{k}|_i^3V}10^{\rho_i},
\end{equation}
with $V$ the surveyed volume in Mpc$^3$, $\Omega_{\mathrm{pix}}$ the image pixel size in radians and the spherical power spectrum coefficients $10^{\rho_i}$ defined in units of mK$^2$ (h$^{-1}$ Mpc)$^3$.

Given these two choices of prior and assuming a uniform prior on the quadratic amplitude parameters $\mathbf{q}$ such that $\mathrm{Pr}(\bmath{q}) = 1$, the conditional distribution  $\mathrm{Pr}(\mathbf{d} | \mathbf{q}, \mathbf{a})$ remains as in Eq. \ref{Eq:DataLike},  while the latter part of Eqn \ref{Eq:Prob} is given by,
\begin{equation}
\label{Eq:ProbFreq}
\mathrm{Pr}(\mathbf{a} | \mathbf{\rho})\mathrm{Pr}(\bmath{\rho}) \; \propto \; \frac{1}{\sqrt{\mathrm{det}\mathbf{\varphi}}} \exp\left[-\mathbf{a}^{*T}\mathbf{\Phi}^{-1}\mathbf{a}\right].
\end{equation}
When assuming a log-uniform prior on the amplitude of the power spectrum coefficients ($\mathrm{Pr}(\bmath{\rho}) = 1$), and when using a prior that is uniform in the amplitude, Eq. \ref{Eq:ProbFreq} becomes,

\begin{equation}
\label{Eq:ProbFreqUni}
\mathrm{Pr}(\mathbf{a} | \mathbf{\rho})\mathrm{Pr}(\bmath{\rho}) \; \propto \; \frac{1}{\sqrt{\mathrm{det}\mathbf{\varphi}}} \exp\left[-\mathbf{a}^{*T}\mathbf{\Phi}^{-1}\mathbf{a}\right]\prod_{s=1}^{N_s}10^{\rho_s}, 
\end{equation}
with $N_s$ the number of spherical power spectrum bins used in the prior.

\subsection{A non-Gaussian prior}
%
%
During the EoR, the emergence of the first stars and galaxies resulted in the gradual formation of ionized `bubbles' in the neutral hydrogen that made up the surrounding intergalactic medium.
The power spectrum of brightness temperature fluctuations in the redshifted 21-cm emission from the EoR describes the magnitude of the 21-cm fluctuations at
different scales. However, this description will be complete only for a Gaussian distribution of 21-cm brightness temperature fluctuations.
While the underlying hydrogen density distribution is expected to be well described as Gaussian after recombination, it develops non-Gaussian features due to the formation of non-linear structures as reionization progresses. Additionally, fluctuations in both the neutral and the ionized hydrogen densities are influenced by the patchiness of reionization (e.g. \citealt{2006MNRAS.372..679M}). As a result, a complete statistical description of the 21-cm brightness temperature distribution must also include higher-order fluctuations.

The approach outlined in Section \ref{Section:IntLike} explicitly assumes that the signal coefficients that fall into a specific $|k|$ bin are well described by a Gaussian random process; however, for a sufficiently high signal-to-noise detection of the EoR signal, a non-Gaussian prior provides a preferred model capable of describing higher order fluctuations present in the brightness distribution. In this paper, we do not consider estimation of a non-Gaussian 21-cm signal. Nevertheless, to aid future work investigating non-Gaussianity in the EoR signal, we describe the necessary modifications to the framework presented here.

To include a non-Gaussian prior, we can use the approach developed in \cite{2001PhRvD..64f3512R}, which is based on the energy eigenmode wavefunctions of a simple harmonic oscillator, and has since been applied to other areas of astrophysics \citep{2014MNRAS.444.3863L}, which we outline in brief below.

For a general random variable $x$, we write the probability density function (PDF) for fluctuations in $x$ as,
\begin{equation}
\label{Eq:PrX}
\mathrm{Pr}(x | \sigma, \bmath{\alpha}) = \exp\left[-\frac{x^2}{2\sigma^2}\right]\left|\sum_{n=0}^{\infty}\alpha_nC_nH_n\left(\frac{x}{\sqrt{2}\sigma}\right)\right|^2 \ ,
\end{equation}
with $\alpha_n$ free parameters that describe the relative contributions of each term to the sum, and $C_n$ is a normalisation factor given by,
\begin{equation}
C_n(\sigma) = \frac{1}{(2^n n! \sqrt{2\pi}\sigma)^{1/2}} \ .
\end{equation}
Equation \ref{Eq:PrX} forms a complete set of PDFs, normalised such that,
\begin{equation}
\int_{-\infty}^\infty \; \mathrm{d}x \; \exp\left[-\frac{x^2}{\sigma^2}\right]C_nH_n\left(\frac{x}{\sqrt{2}\sigma}\right)C_mH_m\left(\frac{x}{\sqrt{2}\sigma}\right) = \delta_{mn},
\end{equation}
with $\delta_{mn}$ the Kronecker delta, where the ground state, $H_0$, reproduces a standard Gaussian PDF, and any non-Gaussianity in the distribution of $x$ will be reflected in non-zero values for the coefficients $\alpha_n$ associated with higher order states.

The only constraint we must place on the values of the amplitudes $\bmath{\alpha}$ is,
\begin{equation}
\sum_{n=0}^{n_{\mathrm{max}}} \left|\alpha_n\right|^2 = 1 \ ,
\end{equation}
with $n_{\mathrm{max}}$ the maximum number of coefficients to be included in the model for the PDF.  This is performed most simply by setting,
 \begin{equation}
\alpha_0 = \sqrt{1 - \sum_{n=1}^{n_{\mathrm{max}}} \left|\alpha_n\right|^2}.
\end{equation}

Using this formalism, we can then parameterise any non-Gaussianity in the coefficients $\mathbf{a}$ by rewriting Eq. \ref{Eq:ProbFreq},
\begin{eqnarray}
\label{Eq:NGRed}
\mathrm{Pr}(\mathbf{a}\; | \;\bmath{\varphi},  \mathbf{\alpha}) &=& \exp\left[-\frac{1}{2}\mathbf{a}^{T}\bmath{\Phi}^{-1}\mathbf{a}\right]\\
 &\times&\prod_{i=1}^{n}\left|\sum_{n=0}^{n_{\mathrm{max}}}\alpha_nC_n(\varphi_{i})H_n\left(\frac{a_i}{\sqrt{2\varphi_{i}}}\right)\right|^2. \nonumber
\end{eqnarray}
The advantage of this method is that one may use a finite set of non-zero $\alpha_n$ to model the non-Gaussianity, without mathematical inconsistency.  Any truncation of the series still yields a proper distribution, in contrast to the more commonly used Edgeworth expansion (e.g. \citealt{2000ApJ...534...25C}).

\subsection{Performing the sampling}

How we now perform the sampling depends entirely on the size of the k-cube we will be using to describe the EoR signal present in the visibilities.  When the size of the k-cube, and thus the number of signal parameters used to describe the signal is small ($< 20000$), we can marginalise over the coefficients $\mathbf{a}$ analytically and sample directly from the power spectrum coefficients $\bmath{\rho}$, a process we describe in Section \ref{Section:Marginalisation}.  In this scenario, we can perform the sampling using {\sc{MultiNest}} \citep{2008MNRAS.384..449F,2009MNRAS.398.1601F}, allowing us to perform robust evidence evaluation, and perform model selection on the EoR power spectrum.

If, however, we wish to sample over a larger number of signal coefficients, the matrix to be inverted when performing the marginalisation analytically will become too large to make this approach computationally tractable\footnote{For HERA, in the limit that the instrumental primary beam and aperture function can be approximated as Gaussian with a Nyquist sampling rate of 4 $uv$-cells area enclosed by the FWHM of the beam and assuming a 38 channel dataset, as used in this paper,  the transition between the small and large $k$-cube regime occurs between the 37-antenna and 61-antenna incremental build-out stages.}.  In this situation we can perform the marginalisation numerically, sampling directly from the high dimension, joint probability distribution described in Eq \ref{Eq:Prob}, a process made possible through the use of a GHS (B18), which we describe in the Section \ref{Section:Sampling}.


\section{The small k-cube regime: Analytical marginalisation over the signal coefficients}
\label{Section:Marginalisation}

In order to perform the marginalisation over the signal coefficients $\mathbf{a}$ and $\mathbf{q}$, we first simplify our notation by defining the vector $\mathbf{b}$, as the concatenation of the vectors $\mathbf{a}$ and $\mathbf{q}$, and the matrix $\mathbfss{T}$, such that our signal can be rewritten,
\begin{equation}
\mathbf{m} = \bar{\mathbfss{F}}\left(\mathbfss{F}_z\bmath{a} + \mathbfss{Q}_z\bmath{q}\right) = \mathbfss{T}\mathbf{b}.
\end{equation}
Introducing the definitions
$\mathbfss{T}^T\mathbfss{N}^{-1}\mathbfss{T} + \mathbf{\Phi}^{-1} \equiv \mathbf{\Sigma}$,
where the elements of the matrix $\mathbf{\Phi}^{-1}$ that correspond to the coefficients $\bmath{q}$ are set to zero and $\mathbfss{T}^T\mathbf{N}^{-1}\mathbf{d} \equiv \mathbf{\bar{d}}$, we can write the log of the joint posterior in Eq \ref{Eq:Prob} as,
%
\begin{equation}
\label{Eq:LogL}
\log \mathrm{Pr}(\bmath{\varphi}, \mathbf{a} , \mathbf{q}\;|\; \mathbf{d}) = -\frac{1}{2} \mathbf{d}^T\mathbfss{T}^T\mathbfss{N}^{-1}\mathbfss{T}\mathbf{d} - \frac{1}{2}\mathbf{b}^T\mathbf{\Sigma}\mathbf{b} + \mathbf{\bar{d}}^T\mathbf{b}.
\end{equation}
Taking the derivative of $\log \mathrm{Pr}(\bmath{\varphi}, \mathbf{a} , \mathbf{q}\;|\; \mathbf{d})$ with respect to $\mathbf{b}$, gives us,
\begin{equation}
\label{Eq:Grada}
\frac{\partial \log \mathrm{Pr}(\bmath{\varphi}, \mathbf{a} , \mathbf{q}\;|\; \mathbf{d})}{\partial \mathbf{b}} =  -\mathbf{\Sigma}\mathbf{b} + \mathbf{\bar{d}}^T \ ,
\end{equation}
which can be solved to give the maximum likelihood vector of coefficients $\hat{\mathbf{b}}$,
\begin{equation}
\label{Eq:amax}
\hat{\mathbf{b}} = \mathbf{\Sigma}^{-1}\mathbf{\bar{d}} \ .
\end{equation}
Re-expressing Eq. \ref{Eq:LogL} in terms of $\hat{\mathbf{b}}$ yields,
\begin{eqnarray}
\log \mathrm{Pr}(\bmath{\varphi}, \mathbf{a} , \mathbf{q}\;|\; \mathbf{d}) &=& -\frac{1}{2} \mathbf{d}^T\mathbfss{T}^T\mathbfss{N}^{-1}\mathbfss{T}\mathbf{d} + \frac{1}{2}\hat{\mathbf{b}}^T\mathbf{\Sigma}\hat{\mathbf{b}} \nonumber \\
& - & \frac{1}{2}(\mathbf{b} - \hat{\mathbf{b}})^T\mathbf{\Sigma}(\mathbf{b} - \hat{\mathbf{b}}) \ .
\end{eqnarray}
The 3rd term in this expression can then be integrated with respect to the $m$ elements in $\mathbf{b}$ to give,
\begin{eqnarray}
I &=& \int_{-\infty}^{+\infty}\mathrm{d}\mathbf{b}\exp\left[-\frac{1}{2}(\mathbf{b} - \hat{\mathbf{b}})^T\mathbf{\Sigma}(\mathbf{b} - \hat{\mathbf{b}})\right] \nonumber \\
&=& (2\pi)^m~\mathrm{det} ~ \mathbf{\Sigma}^{-\frac{1}{2}}.
\end{eqnarray}
Our marginalised probability distribution for a set of EoR power spectrum coefficients is then given as,
\begin{eqnarray}
\label{Eq:Margin}
\mathrm{Pr}(\mathbf{\varphi} \;|\; \mathbf{d}) &\propto& \frac{\mathrm{det} \left(\mathbf{\Sigma}\right)^{-\frac{1}{2}}}{\sqrt{\mathrm{det} \left(\mathbf{\varphi}\right)~\mathrm{det}\left(\mathbf{N}\right)}} \\
&\times&\exp\left[-\frac{1}{2}\left(\mathbf{d}^T\mathbf{N}^{-1} \mathbf{d} - \mathbf{\bar{d}}^T\mathbf{\Sigma}^{-1}\mathbf{\bar{d}}\right)\right]. \nonumber
\end{eqnarray}
When taking this approach in Section \ref{Section:Sims} we use the MAGMA (Matrix Algebra on GPU and Multicore Architectures) GPU accelerated linear algebra package\footnote{http://icl.cs.utk.edu/magma/} to perform the Cholesky decomposition for each likelihood evaluation.

\section{The large k-cube regime: Numerical marginalisation over the signal coefficients}
\label{Section:Sampling} 

For a detailed account of both Hamiltonian Monte Carlo (HMC) and GHS refer to B18 or \cite{2013PhRvD..87j4021L}; here we will provide only a brief introduction of the key aspects of each.

HMC sampling \citep{1987PhLB..195..216D}  has been widely applied in Bayesian computation \citep{Neal1993}, and has been successfully applied to problems with extremely large numbers of dimensions ($\sim 10^6$ see e.g. \citealt{2008MNRAS.389.1284T}).  Where conventional MCMC methods move through the parameter space by a random walk and, therefore, require a prohibitive number of samples to explore-high dimensional spaces, HMC exploits techniques that describe the motion of particles in potential wells and suppresses this random walk behaviour. This allows the HMC approach to maintain a reasonable efficiency, even for high-dimensional problems.

Possibly the main shortcoming of traditional HMC methods is that it requires a large number of tuning parameters in order to navigate the parameter space.  In particular, every parameter requires a step size and the total number of steps in each iteration of the sampler must also be chosen.  These are typically determined via expensive tuning runs.  The GHS is designed to bypass much of this tuning by using the Hessian of the sampled probability distribution, calculated at its peak, to set the step size and covariance of the parameter space.   The number of steps at each iteration is then drawn from a uniform distribution U(1, nmax), with nmax of ten found to be suitable for all tested problems. A single global scaling parameter for the step size is then the only tunable parameter, chosen such that the acceptance rate for the GHS is $\sim$68\%. 

Defining the ``potential energy'' $\Psi$ as,
\begin{equation}
\Psi = -\log \mathrm{Pr}(\bmath{\varphi}, \mathbf{a} , \mathbf{q}\;|\; \mathbf{d}) \ ,
\end{equation}
in order to perform sampling we need the following:

\begin{itemize}
\item the gradient of $\Psi$ for each parameter $x_i$,
\item the peak of the joint distribution,
\item the Hessian at that peak.
\end{itemize}
The gradients of our parameters are given by the following:
\begin{equation}
\frac{\partial \Psi}{\partial \mathbf{b}} = -(\mathbf{d} - \mathbfss{T}\mathbf{b})^T\mathbf{N}^{-1}\mathbfss{T} + \mathbf{b}^T\mathbf{\Phi}^{-1}
\end{equation}
\begin{equation}
\label{Eq:rhoderiv}
\frac{\partial \Psi}{\partial \rho_i} = \frac{1}{2}\mathrm{Tr}\left(\mathbf{\Phi}^{-1} \frac{\partial \mathbf{\Phi}}{\partial \rho_i} \right) - \frac{1}{2}\mathbf{b}^T\mathbf{\Phi}^{-1}\frac{\partial \mathbf{\Phi}}{\partial \rho_i}\mathbf{\Phi}^{-1}\mathbf{b}
\end{equation}
and the components of the Hessian are,

\begin{equation}
\frac{\partial^2 \Psi}{\partial \mathbf{b}^2} = \mathbfss{T}^T\mathbf{N}^{-1}\mathbfss{T} + \mathbf{\Phi}^{-1}
\end{equation}

\begin{equation}
\label{Eq:rhodderiv}
\frac{\partial^2 \Psi}{\partial \rho_i^2} = \mathbf{b}^{T}\mathbf{\Phi}^{-1}\frac{\partial \mathbf{\Phi}}{\partial \rho_i}\mathbf{\Phi}^{-1}\frac{\partial \mathbf{\Phi}}{\partial \rho_i}\mathbf{\Phi}^{-1}\mathbf{b}  - \frac{1}{2}\mathbf{b}^{T}\mathbf{\Phi}^{-1}\frac{\partial^2 \mathbf{\Phi}}{\partial \rho_i^2}\mathbf{\Phi}^{-1}\mathbf{b}
\end{equation}

\begin{equation}
\frac{\partial^2 \Psi}{\partial\rho_i\partial \mathbf{b}} = - \mathbf{\Phi}^{-1}\frac{\partial \mathbf{\Phi}}{\partial\rho_i}\mathbf{\Phi}^{-1}\mathbf{b} \ .
\end{equation}

For a set of power spectrum coefficients $\mathbf{\rho}$, we can solve for the maximum set of signal coefficients $\mathbf{b}_{\mathrm{max}}$ analytically using Eq. \ref{Eq:amax}; so, when searching for the global maximum, we need only search over the subset of parameters $\mathbf{\rho}$.  This is achieved by using either a particle swarm algorithm (\citealt{Kennedy1,Kennedy2}; for uses in cosmological parameter estimation see e.g. \citealt{2012PhRvD..85l3008P}) or gradient search optimization \citep{Gilbert1989}.

\subsection{Low signal-to-noise parameterisation}

In \cite{2016arXiv161205258L}, an alternative parameterisation of the likelihood described in Eq.~\ref{Eq:ProbFreq} is described that is much more efficient in the low signal-to-noise regime, where the power spectrum coefficients are not detected.  We can anticipate that, at least at first, this is likely to be the case with the EoR signal, and thus we summarise this new parameterisation in the context of our three-dimensional power spectrum analysis below.

Rather than sample from the parameters $\mathbf{a}$, we instead sample from the related parameters $\mathbf{u}$, where for the $i$th signal amplitude we will have,
\begin{equation}
a_i = u_i\sqrt{\varphi_i},
\end{equation}
where as before $\varphi_i$ is the three-dimensional power spectrum coefficient that describes the standard deviation of the $i$th amplitude parameter. In order to still sample uniformly in the original parameters, $\mathbf{a}$, we then include an additional term, the determinant of the Jacobian describing the transformation from $a_i$ to $u_i$.  The Jacobian in this case has elements,

\begin{equation}
J_{i,j} = \sqrt{\varphi_i}\delta_{i,j,},
\end{equation}
with $\delta_{i,j,}$ the Kronecker delta.  The determinant is therefore,

\begin{equation}
\mathrm{det}\left(\mathbfss{J}\right) = \prod_{i=0}^{m}\sqrt{\varphi_i},
\end{equation}
which acts to cancel exactly with the determinant of the matrix $\mathbf{\Psi}$ in Eq.~\ref{Eq:ProbFreq}. In the following work, when using the GHS, we will use  both parameterisations dependent upon whether we are in the high or low signal-to-noise regime.

\section{Application to simulations}
\label{Section:Sims}

\begin{figure*}
\begin{center}$
\begin{array}{ccc}
\hspace{-1.0cm}
\includegraphics[width=60mm]{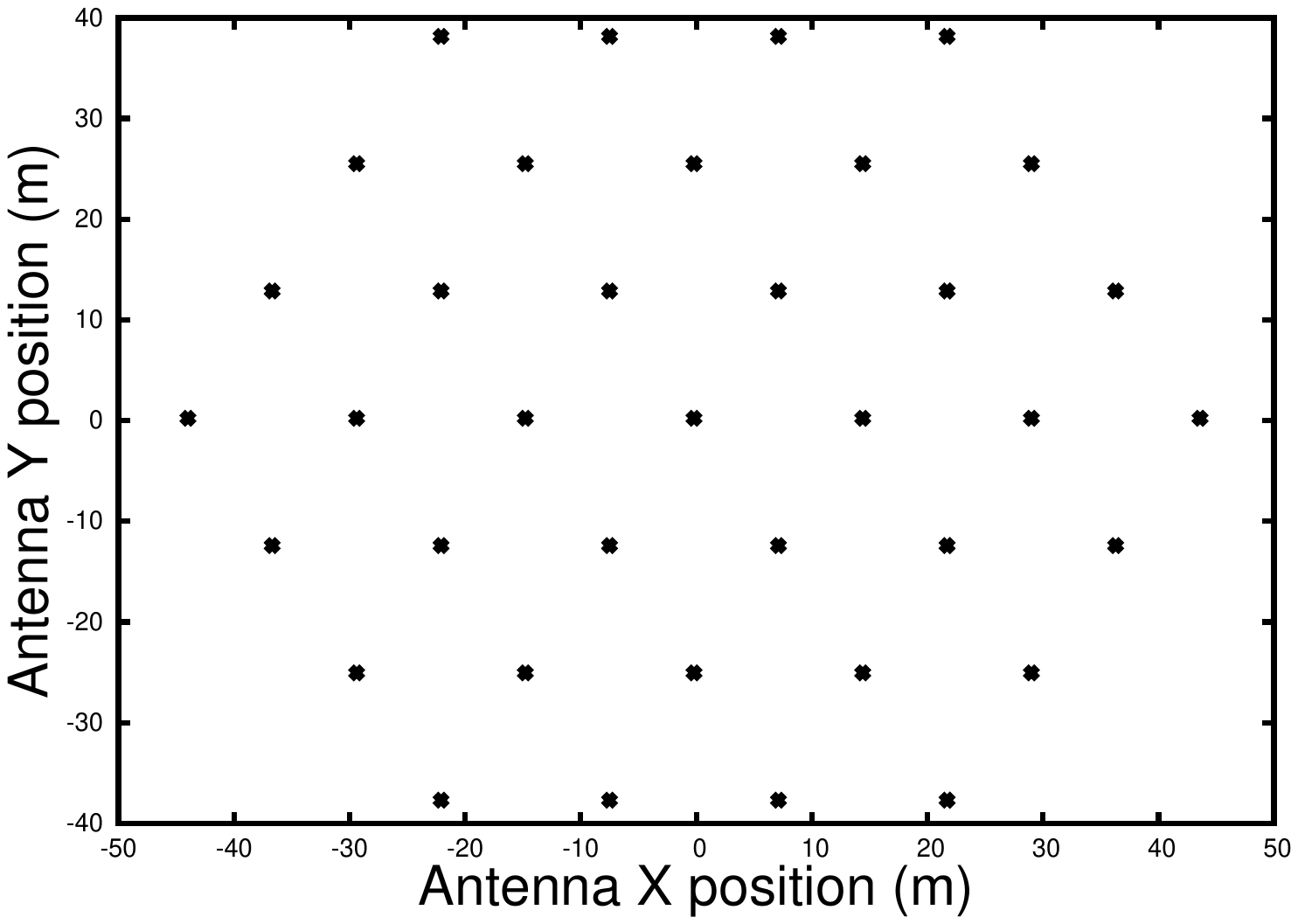} &
\includegraphics[width=60mm]{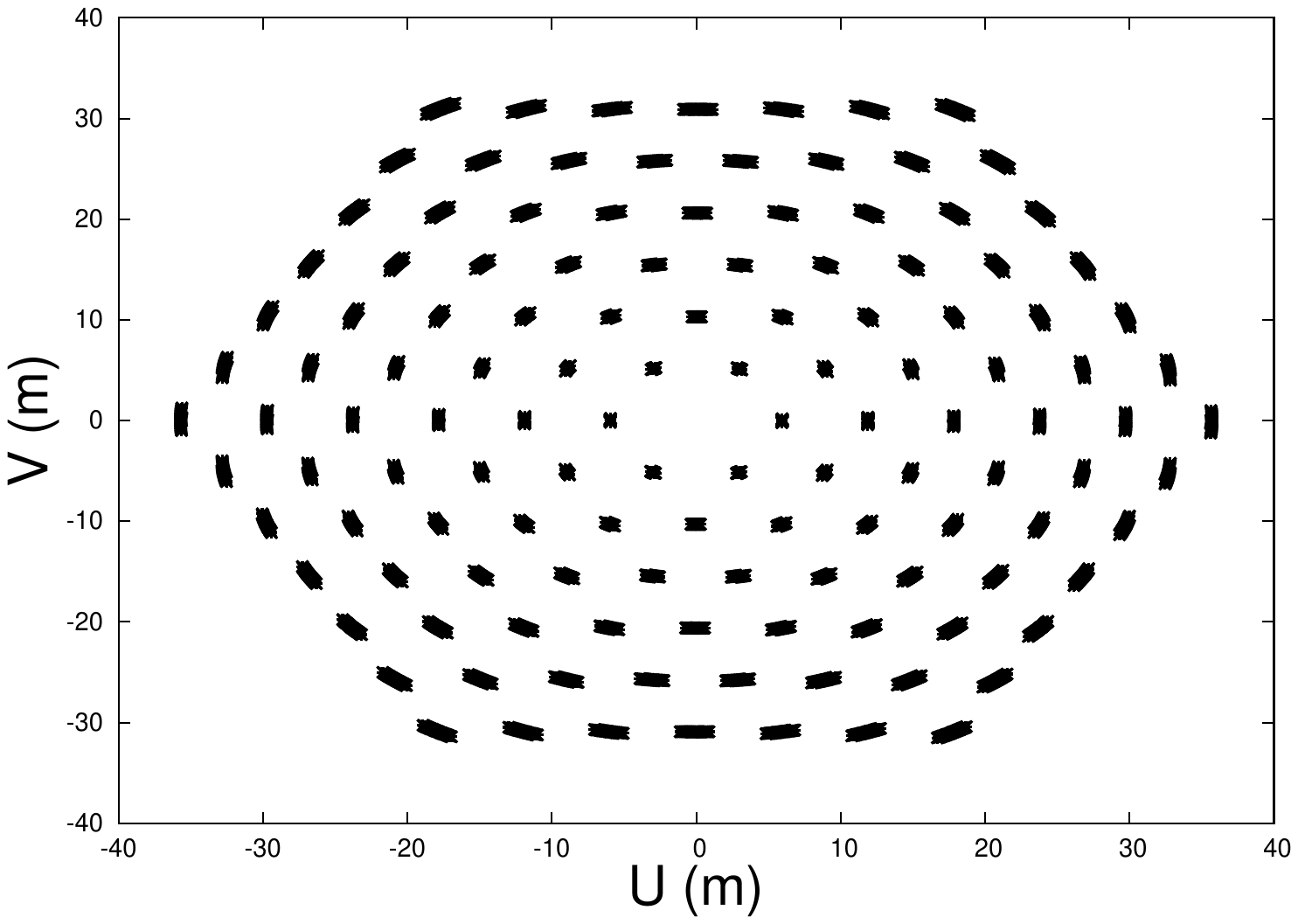} &
\hspace{-1.0cm}
\includegraphics[width=80mm]{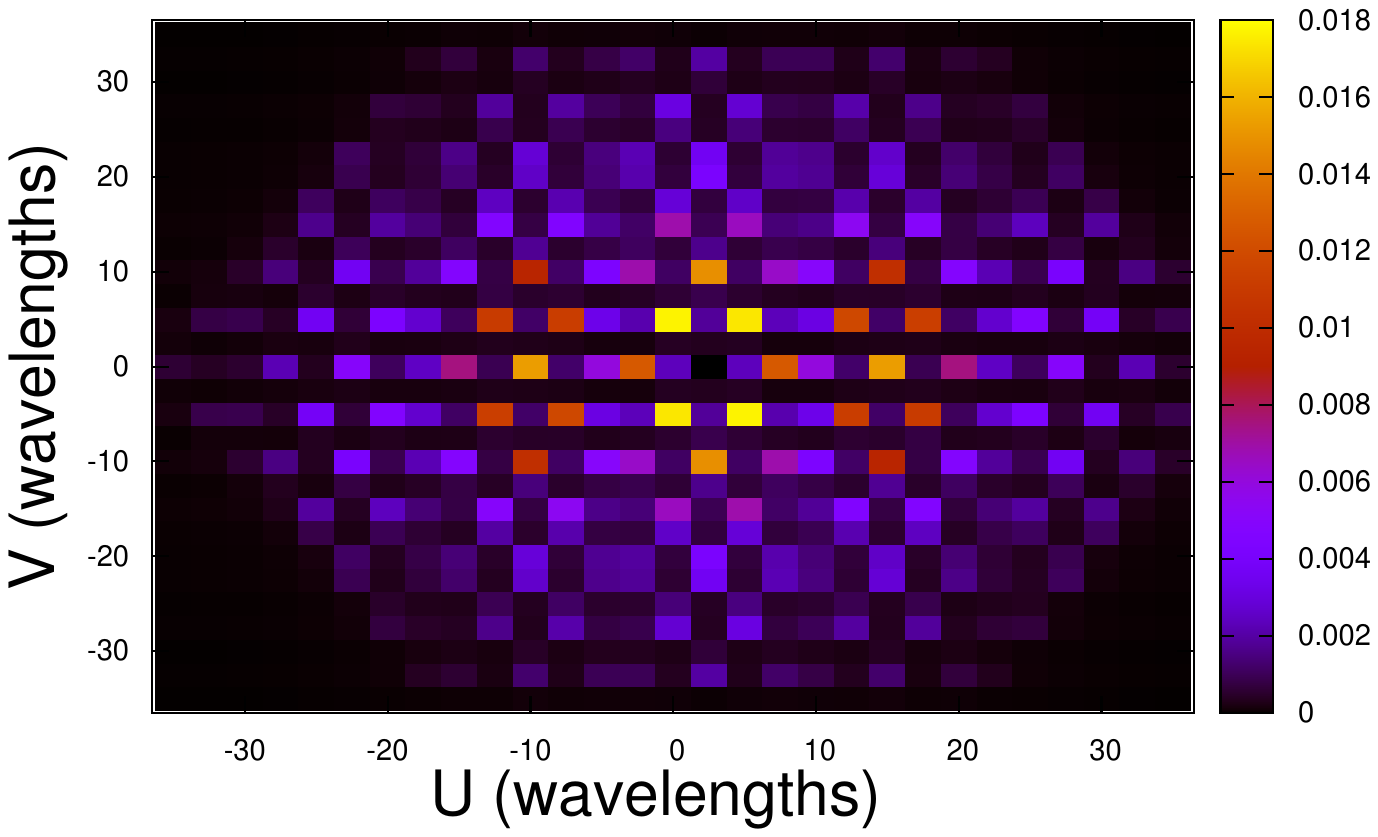} \\
\hspace{-1.0cm}
\includegraphics[width=60mm]{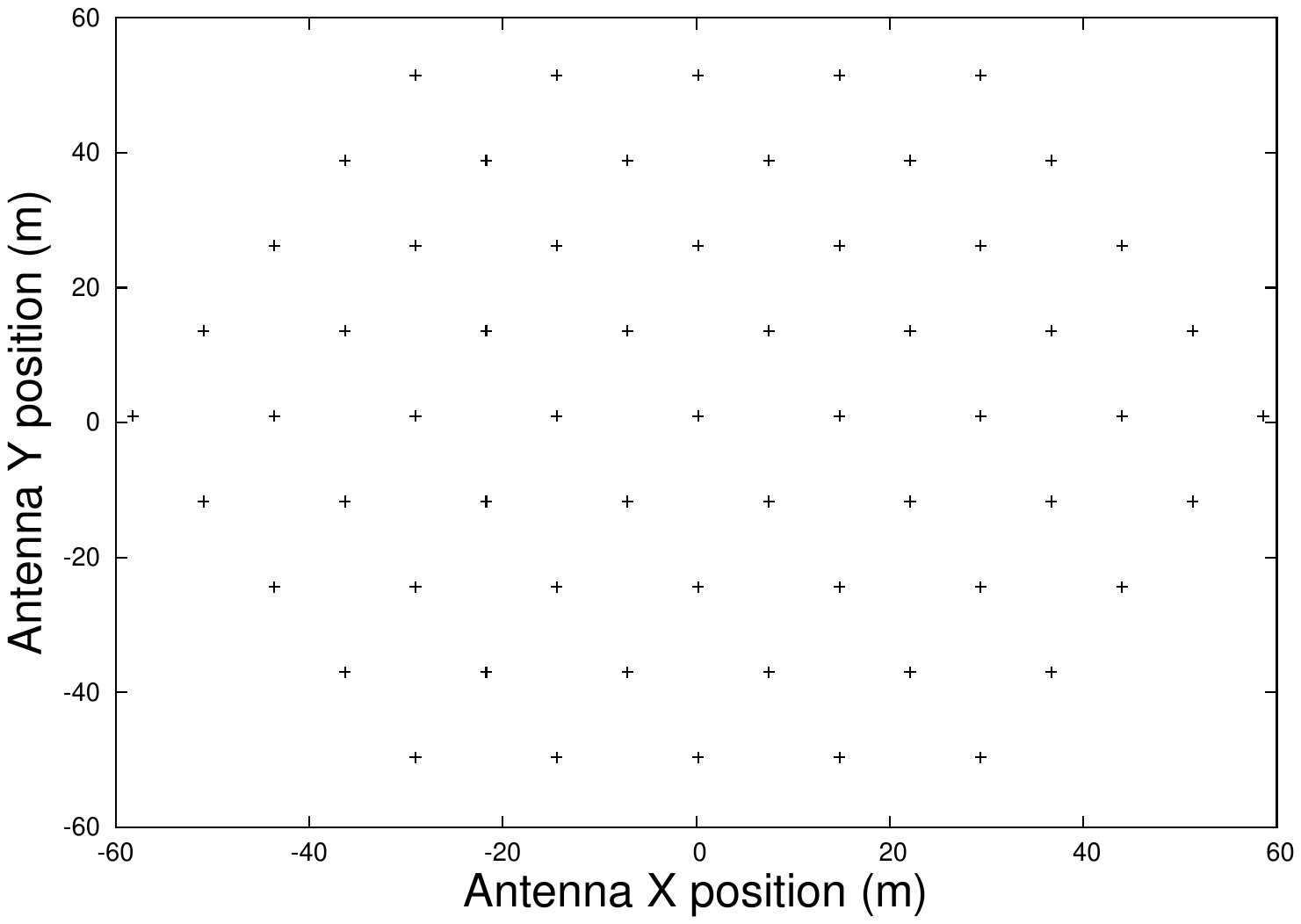} &
\includegraphics[width=60mm]{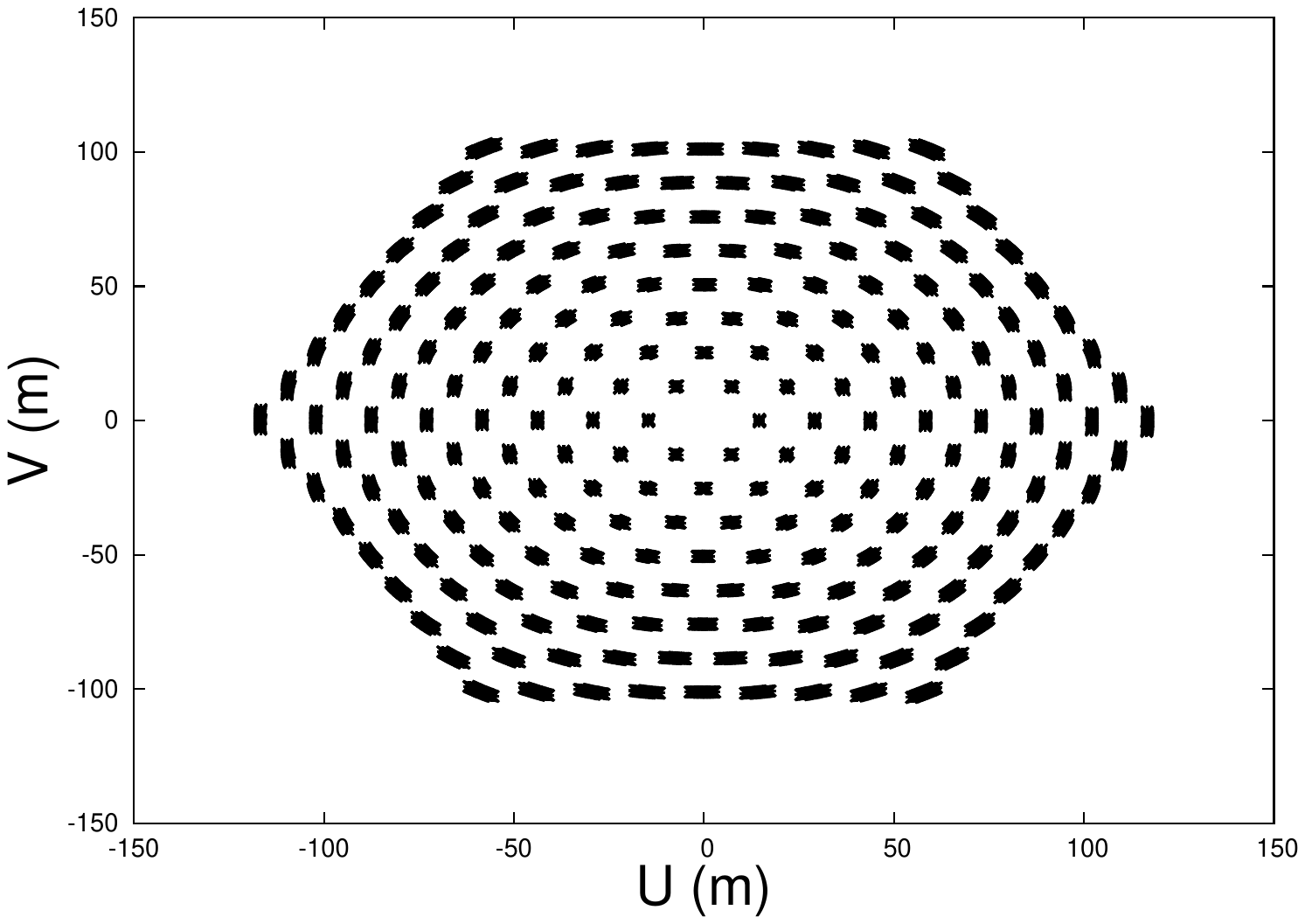} &
\hspace{-1.0cm}
\includegraphics[width=80mm]{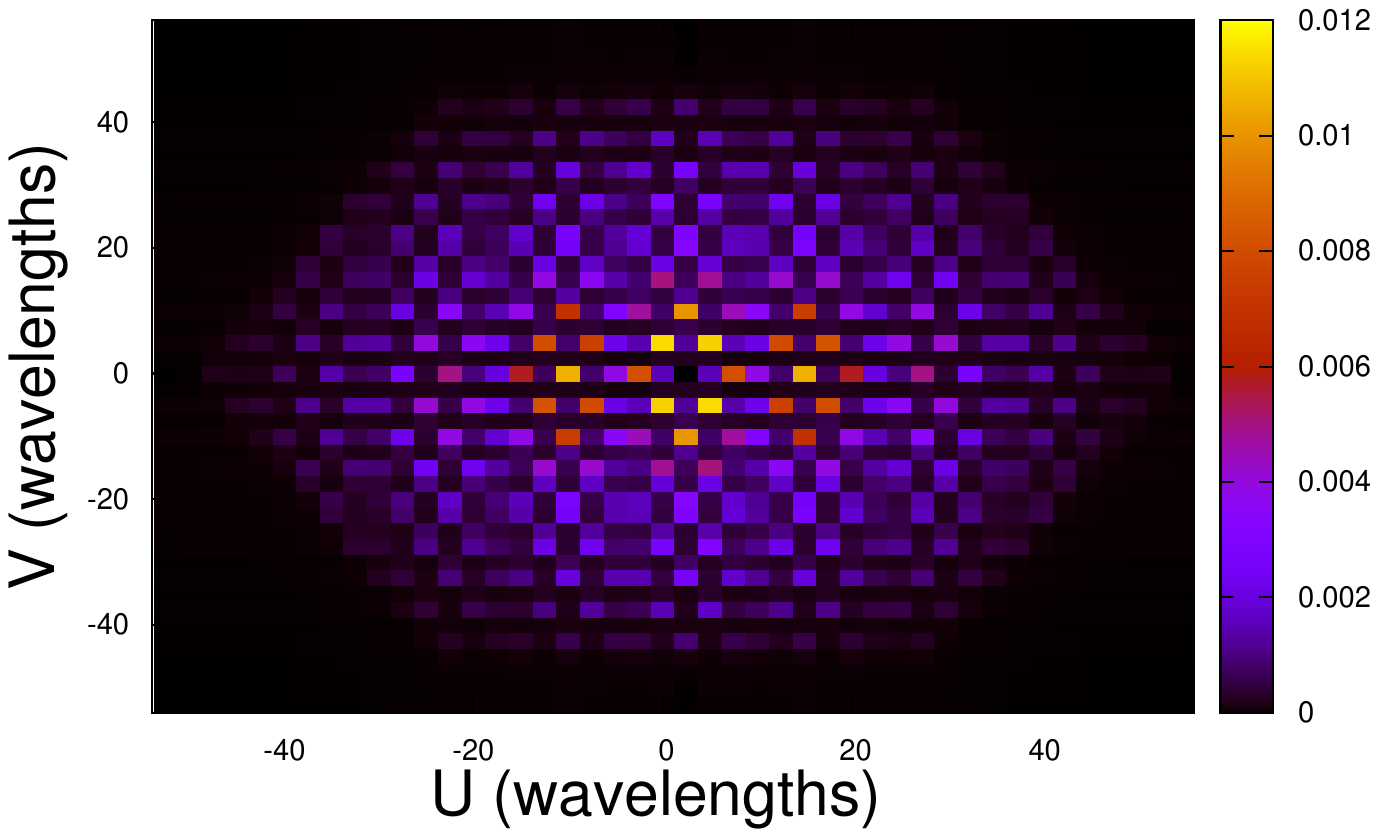} \\
\end{array}$
\end{center}

\caption{[Top] (Left) Antenna positions for the 37 element interferometer used in simulations 1-3 in Section \ref{Section:Sims}.  (Middle) Sampled $(u,v)$ coordinates from a 30 minute observation, and (Right) the relative weights of the $(u,v)$ obtained from Eq. \ref{Eq:Weight}, normalised to have a sum of 1. [Bottom] As [Top], but for the 61 element array used in simulations 4-5 in Section \ref{Section:Sims}.\label{figure:SimsSetup}}
\end{figure*}

\begin{figure*}
\begin{center}$
\begin{array}{cc}
\hspace{-1cm}
\includegraphics[width=100mm]{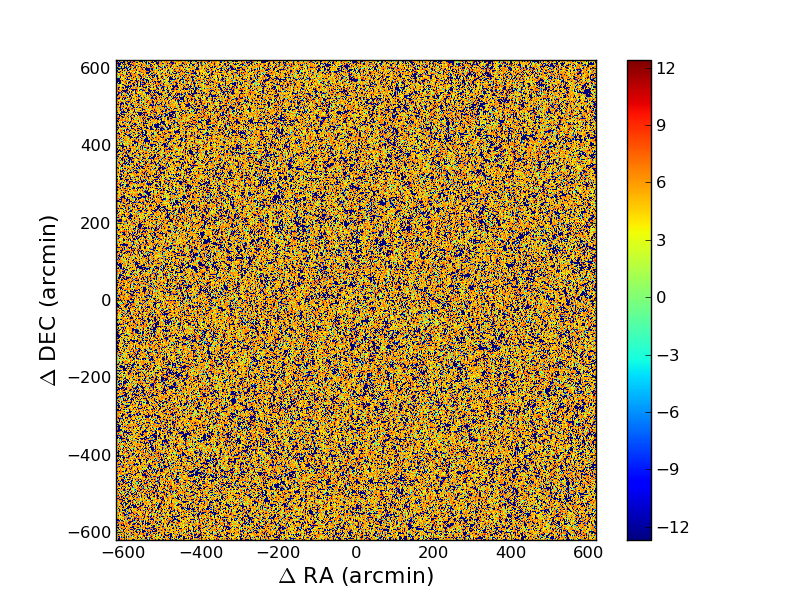} &
\hspace{-1cm}
\includegraphics[width=100mm]{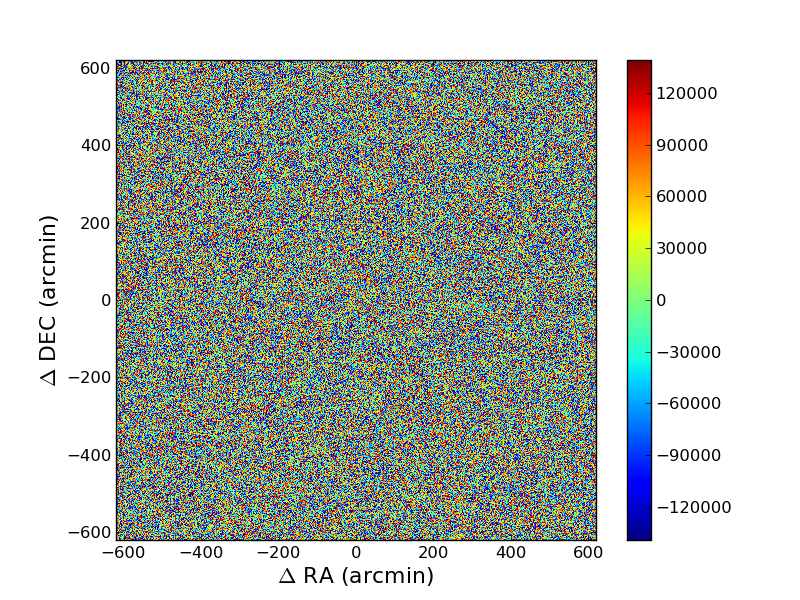}  \\
\end{array}$
\end{center}

\caption{(left) A single channel from the simulation of the EoR signal using the seminumerical 21cmFAST algorithm \citep{2011MNRAS.411..955M,2007ApJ...669..663M}  described in Section \ref{Section:Sims}, after subtracting the mean in the spatial variations for that channel.  (right) A single channel from a simple continuum sky simulation described in Section \ref{Section:Sims}, after subtracting the mean in the spatial variations for that channel.  The continuum simulation is scaled such that the power is a factor $10^8$ greater than in the EoR simulation. The channels are displayed at the $0.7$ arcmin resolution at which we simulate the EoR and foreground image cubes.
In both cases, the colour scale is in mK.}
\label{figure:ChanSims}
\end{figure*}


We now apply the methods described in the preceding sections to a series of five simulations.  In the first three of these we perform the simulation using a 37 element interferometer, with antenna locations shown in Fig. \ref{figure:SimsSetup} (top, left), while for the final two simulations we use a 61 element array (Fig. \ref{figure:SimsSetup} bottom, left).  These antenna configurations are representative of the HERA 37 and 61 element arrays.  In both cases, we simulate 38 200\,kHz channels spanning the range 122.17$\sim$129.90\,MHz. EoR instruments typically employ significantly larger instantaneous bandwidths (50--250 MHz, in the case of HERA; \citealt{2017PASP..129d5001D}); however, cosmological evolution of the 21-cm signal as a function of redshift limits the bandwidth that can be used in a power spectrum measurement to $\sim 8\,$MHz \citep{2006PhR...433..181F}.

\begin{table}
\centerline{
\begin{tabular}{l l l}
\hline
Parameter                & Description              & Value  \\
\hline
$\eta_{s}$               & System efficiency        & 1    \\
$\Delta\nu$              & Frequency channel width  & $200~\mathrm{kHz}$     \\
$\eta_{a}$               & Antenna efficiency       & 1    \\
$A$                      & Antenna effective area   & $150~\mathrm{m^2}$\\
$T_{\mathrm{sys}}$	 & System temperature	    & $550~\mathrm{K}$\\	
\hline
\end{tabular}
}
\caption{Instrument and observation parameters used to calculate the variance of the theoretical instrumental thermal noise per unit time $\tau$.}
\label{Table:instrument}
\end{table}

We calculate the visibility domain theoretical instrumental thermal noise for our simulation per unit time $\tau$, given the parameters in Table \ref{Table:instrument}, as in \cite{1999ASPC..180.....T},
\begin{equation}
\sigma(\tau) = 10^{-26}\frac{1}{\eta_{\mathrm{s}}}\frac{2k_bT_{\mathrm{sys}}}{\eta_{\mathrm{a}}A}\frac{1}{\sqrt{2\Delta\nu\tau}}\, \mathrm{Jy} \ .
\end{equation}
For each simulation, we use the array configurations for the 37 or 61 element interferometers in Fig.\ref{figure:SimsSetup} as input to the {\sc{CASA}}\footnote{http://casa.nrao.edu} (Common Astronomy Software Applications) \texttt{simobserve} tool, to obtain the set of observed $(u,v)$ coordinates that correspond to a series of 30 second integrations over a single 30 minute pointing, given those configurations. We take the pointing centre to have right ascension equal to 0\fdg0, and declination equal to -30\fdg0.  This results in 21870 sampled $(u,v)$ coordinates per channel for the 37 element array and 58141 per channel for the 61 element array (Fig. \ref{figure:SimsSetup}, middle, top and bottom panels, respectively).

Our input sky models are constructed using $2048\times 2048$ pixels, with a resolution of $\sim$ 40 arcseconds per pixel, giving a total field of view of $\sim 23\times23$ degrees.  We then multiply these sky models by a Gaussian primary beam with a full width at half max of 8 degrees at 122.17~MHz, and evaluate the direct Fourier transform of the observed sky-models onto the sampled $(u,v)$ points obtained previously. 

We now describe each of the five simulations in more detail below:

\begin{description}
  \item[Simulation 1] \hfill \\
  A 2000 hrs simulation of a single flat spectrum point source, 10.4 degrees away from the primary beam center, resulting in a 1000$\sigma$ detection using the 37 element array shown in Fig. \ref{figure:SimsSetup} (top).  We include uncorrelated thermal noise in each visibility.  To simulate 4000 repetitions of our 30 minute observation, we therefore add noise with an rms of 0.045\,Jy to each of the 21870 visibilities in each channel. In order to compare equivalent simulations we use the same white noise realization for simulations 1-3, and for simulations 4-5.  For ease of interpretation we do not include the quadratic described in Eq. \ref{Eq:Quad} in our model when analysing this simulation.  The purpose of this first simulation is to show in a straightforward way how our approach automatically accounts for the frequency dependence of the UV-sampling, which causes observed low frequency structure along individual baselines.\\

  \item[Simulation 2] \hfill \\
  A 160 hour integration including only the EoR signal and the uncorrelated thermal noise described in Simulation 1.  We generate the EoR signal using the seminumerical 21cmFAST algorithm \citep{2011MNRAS.411..955M,2007ApJ...669..663M} to simulate a cosmological volume of $1024^3$~Mpc$^3$.  We use this same EoR realisation in all subsequent simulations. An example of one channel from this EoR simulation is shown in Fig. \ref{figure:ChanSims} (left). \\
  
  \item[Simulation 3] \hfill \\
  As Simulation 2, but with an additional flat spectrum continuum component added to the model, shown in Fig. \ref{figure:ChanSims} (right).  Each pixel in the continuum model is assigned a random positive value, drawn uniformly between zero and one, which is then held constant across the 38 channels for that pixel.  We then scale the image so that the total power in the mean subtracted continuum is $\sim 10^8$ times that of the EoR signal. An example of one channel from this continuum simulation is shown in Fig. \ref{figure:ChanSims} (right).  \\
  
  \item[Simulation 4] \hfill \\
  As Simulation 2, however we use the 61 element array shown in Fig. \ref{figure:SimsSetup} (bottom). \\
  
  \item[Simulation 5] \hfill \\
  As Simulation 4, but an additional flat spectrum continuum component is added to the model, as described in Simulation 3.\\
\end{description}

In order to adequately sample the aperture function of the Gaussian primary beam in the UV plane, we define our UV cells to each have a width of 2.5$\lambda$. We then use Eq. \ref{Eq:Weight} to determine the set of cells to include in our model.  The weights for each cell are shown in Fig. \ref{figure:SimsSetup} (right) for the 37 and 61 element arrays (top and bottom panels, respectively).  Including all cells that contribute up to 99$\%$ of the total weight, we find results in 650 and 1142 UV cells per $\eta$ mode included in the model for the 37 and 61 element arrays, respectively.

For simulations 1-3 we will use the analytic marginalisation over the signal coefficients described in Section \ref{Section:Marginalisation}, sampling from the 7 dimensional spherical power spectrum using the {\sc{MultiNest}} algorithm.  For simulations 4-5 the number of signal coefficients included in the model is too great for this analytic approach, and so we perform this marginalisation numerically using the GHS described in Section \ref{Section:Sampling}.


%
%

\subsection{Results for Simulation 1}

\begin{figure*}
\begin{center}$
\begin{array}{cc}
\hspace{-1cm}
\includegraphics[width=80mm]{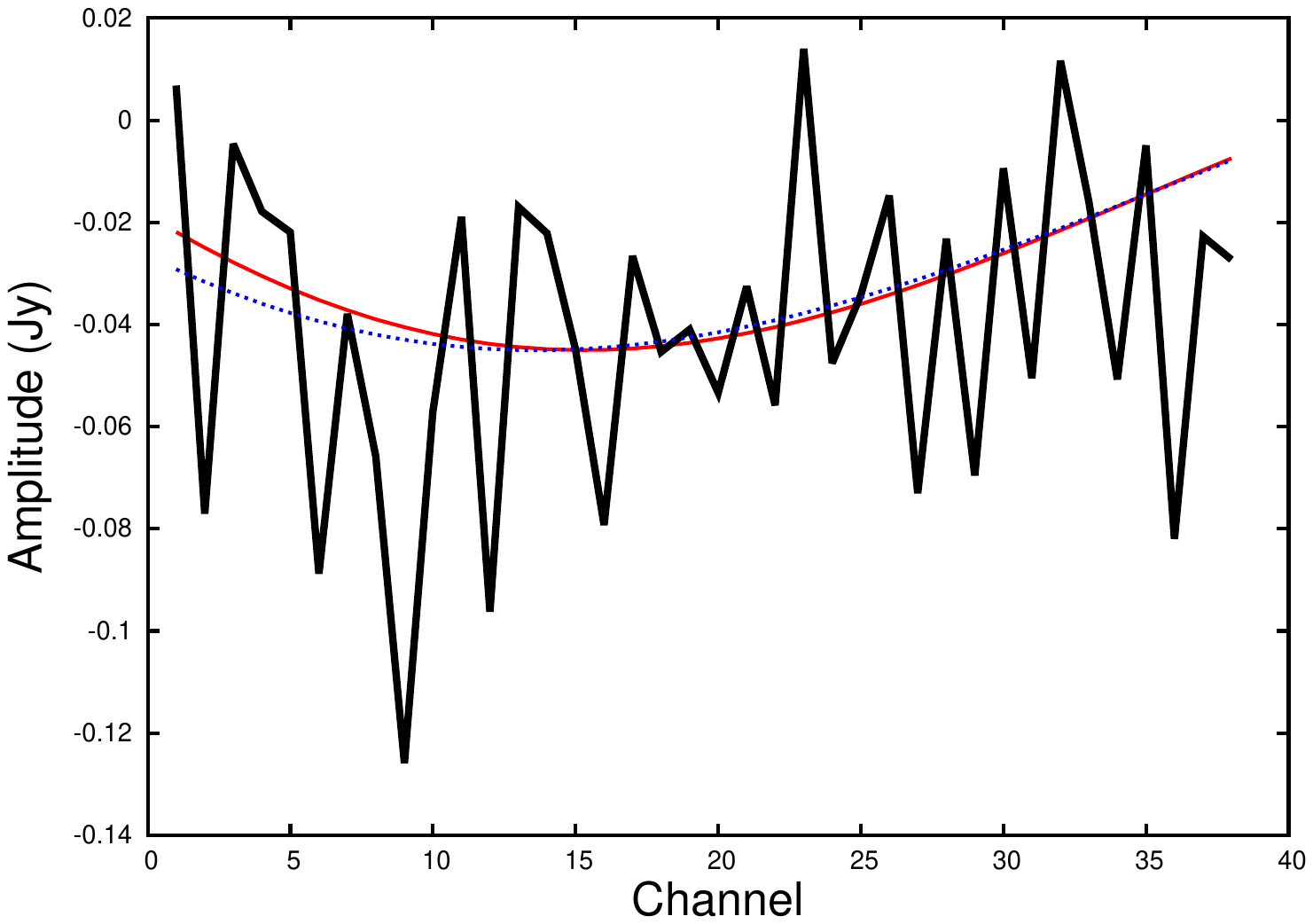} &
\includegraphics[width=80mm]{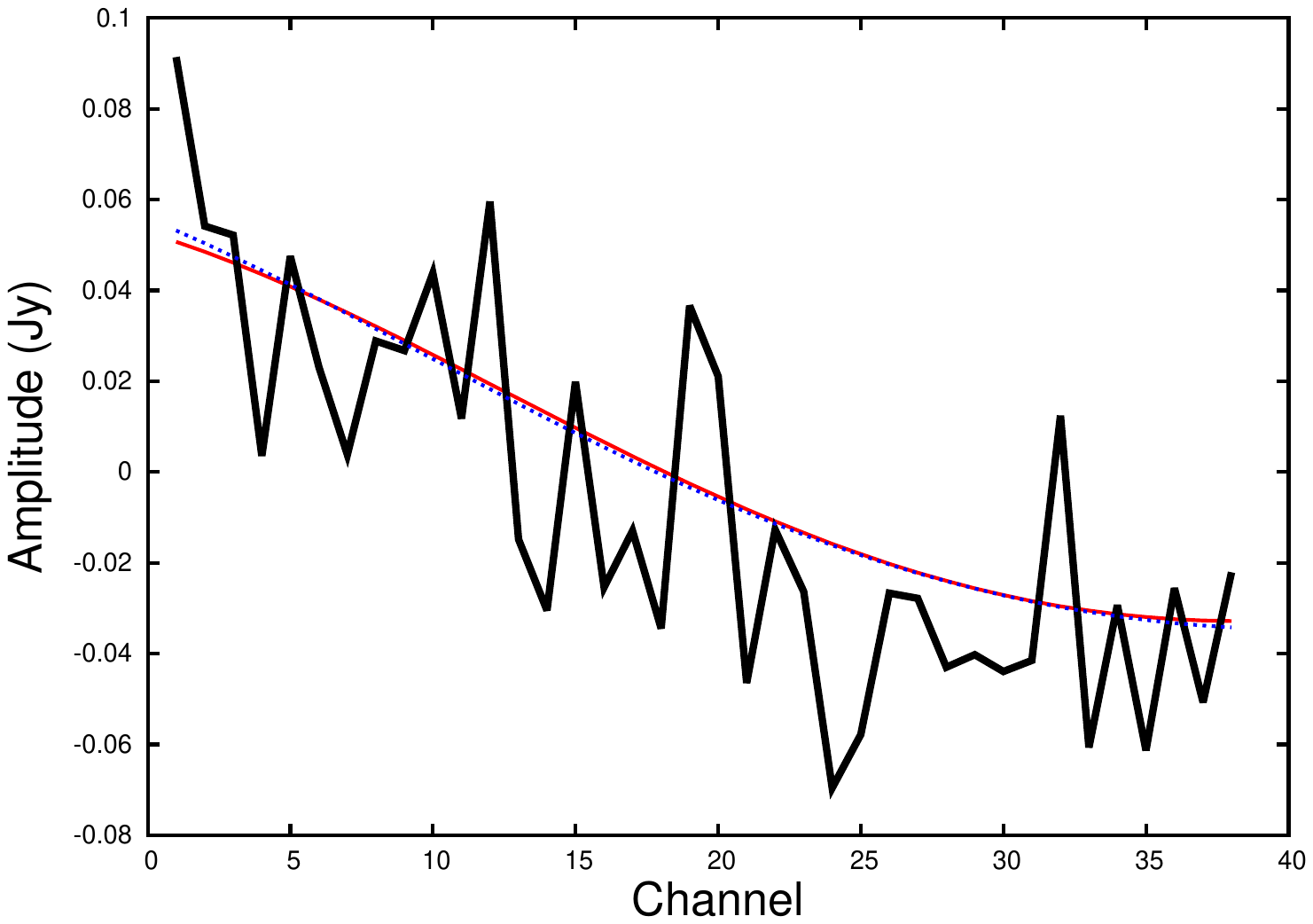} \\
\includegraphics[width=80mm]{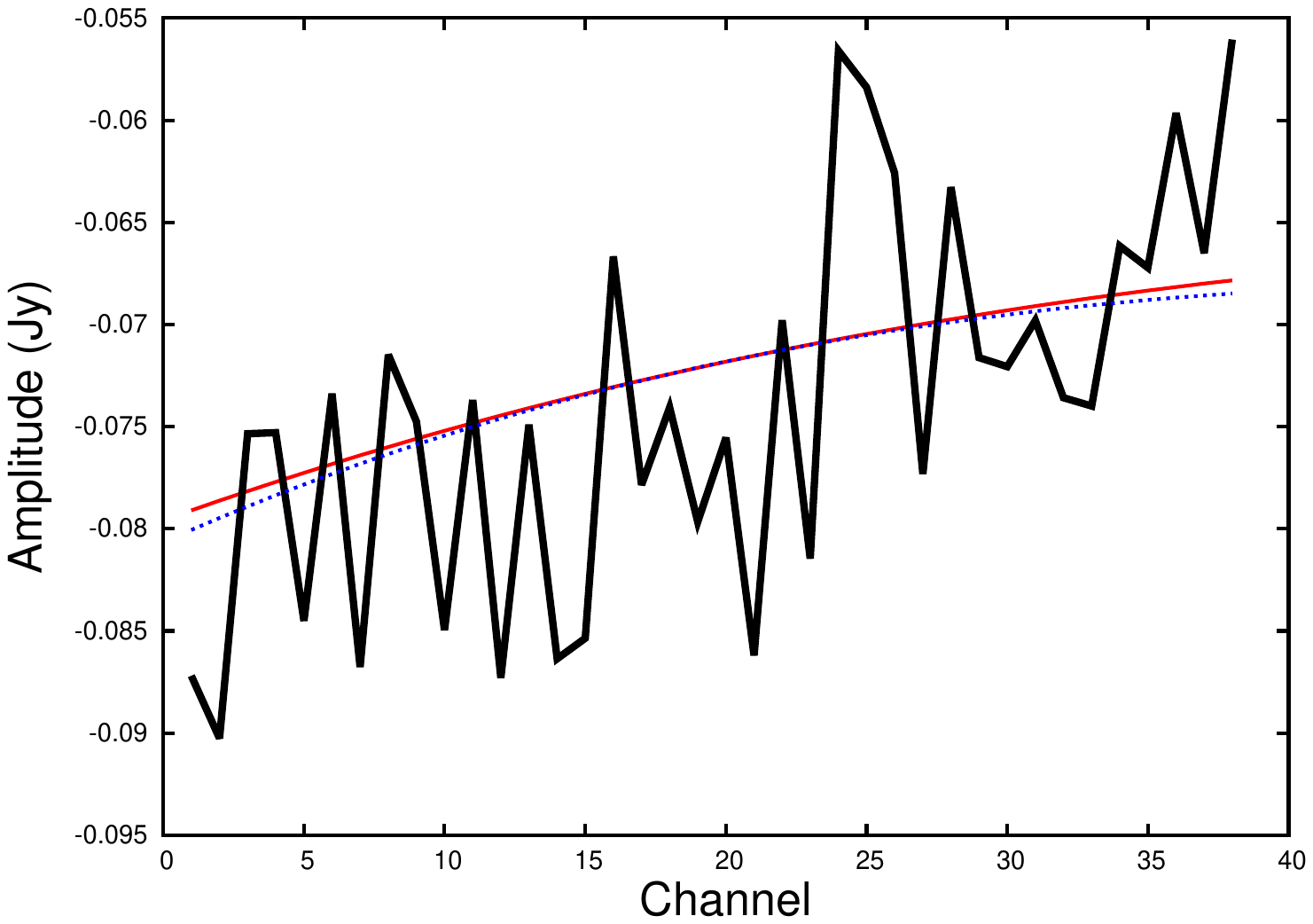} &
\includegraphics[width=80mm]{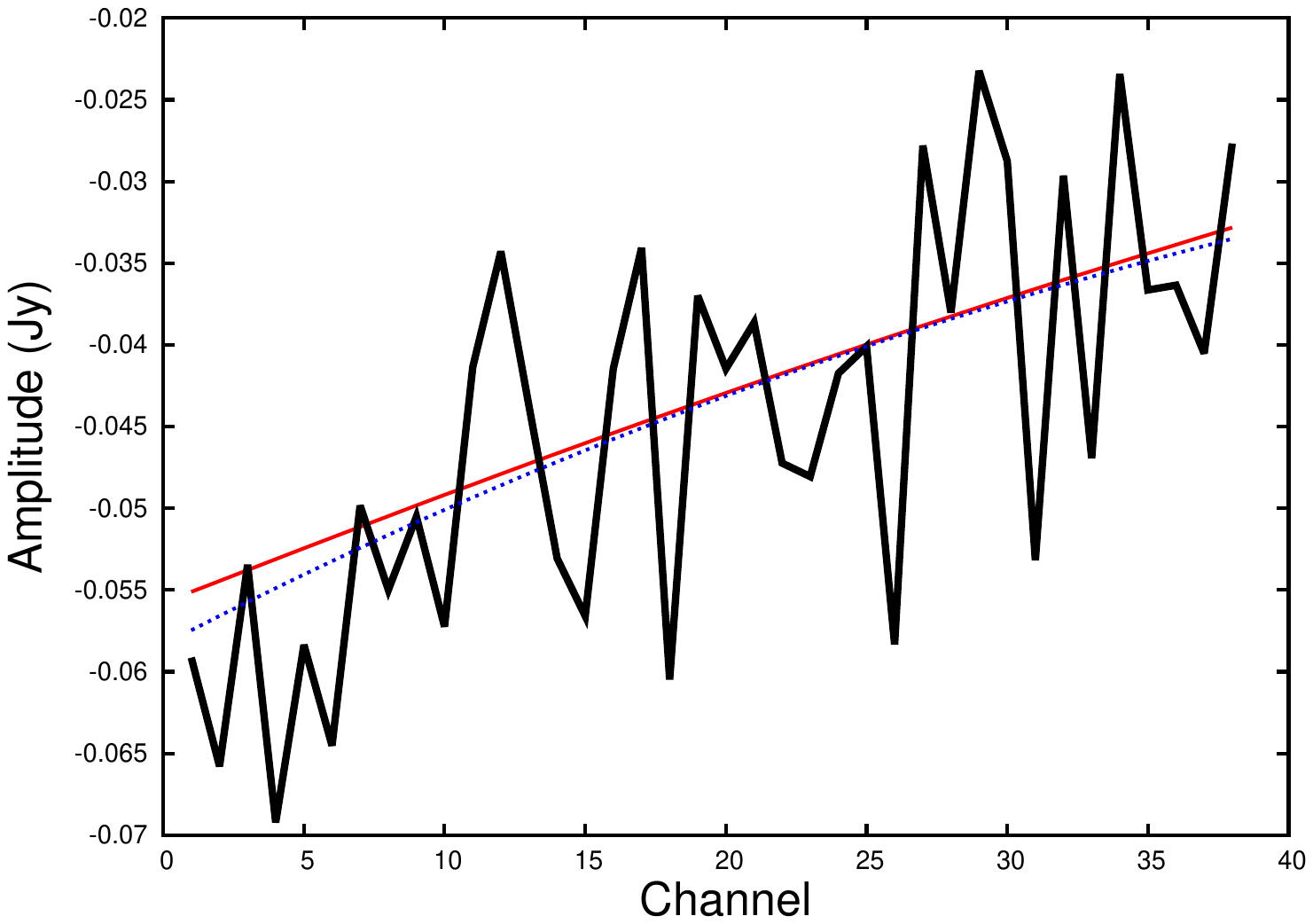} \\
\end{array}$
\end{center}

\caption{The real (left) and imaginary (right) components from the longest (top) and shortest (bottom) baselines taken from simulation 1. The red line in each case is the injected data, the black line is the injected data with added uncorrelated noise, and the blue line is the maximum likelihood recovered signal from our analysis for the offset term in our model.  This simulation contains a single flat spectrum point source, however, as each baseline samples a range of different $(u,v)$ coordinates as a function of frequency, it is seen to have structure. We note here that we did not include the quadratic described in Eq.\ref{Eq:Quad} in our model for this simulation, the recovered structure comes solely as a result of projecting our model k-cube onto the sampled visibility points. \label{figure:Baselines}}
\end{figure*}

\begin{figure*}
\begin{center}$
\begin{array}{ccc}
\hspace{-1.3cm}
\includegraphics[width=80mm]{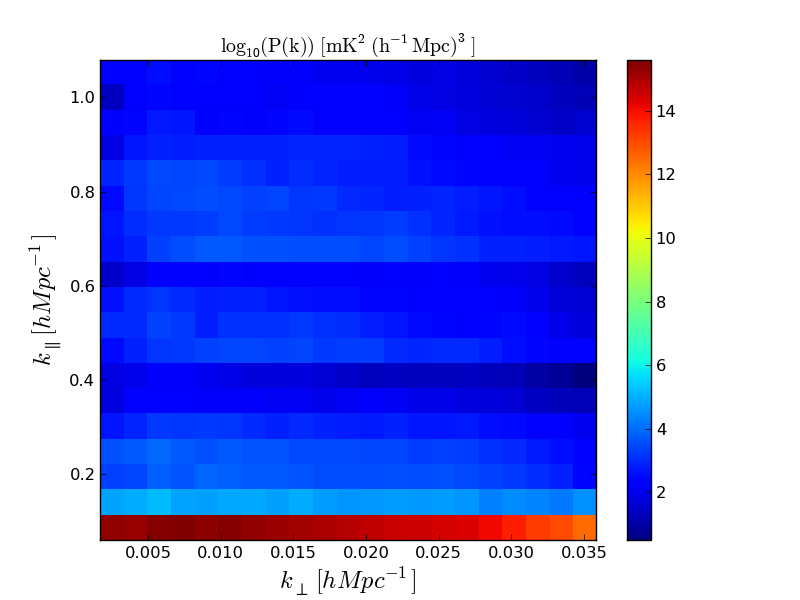} &
\hspace{-1.2cm}
\includegraphics[width=80mm]{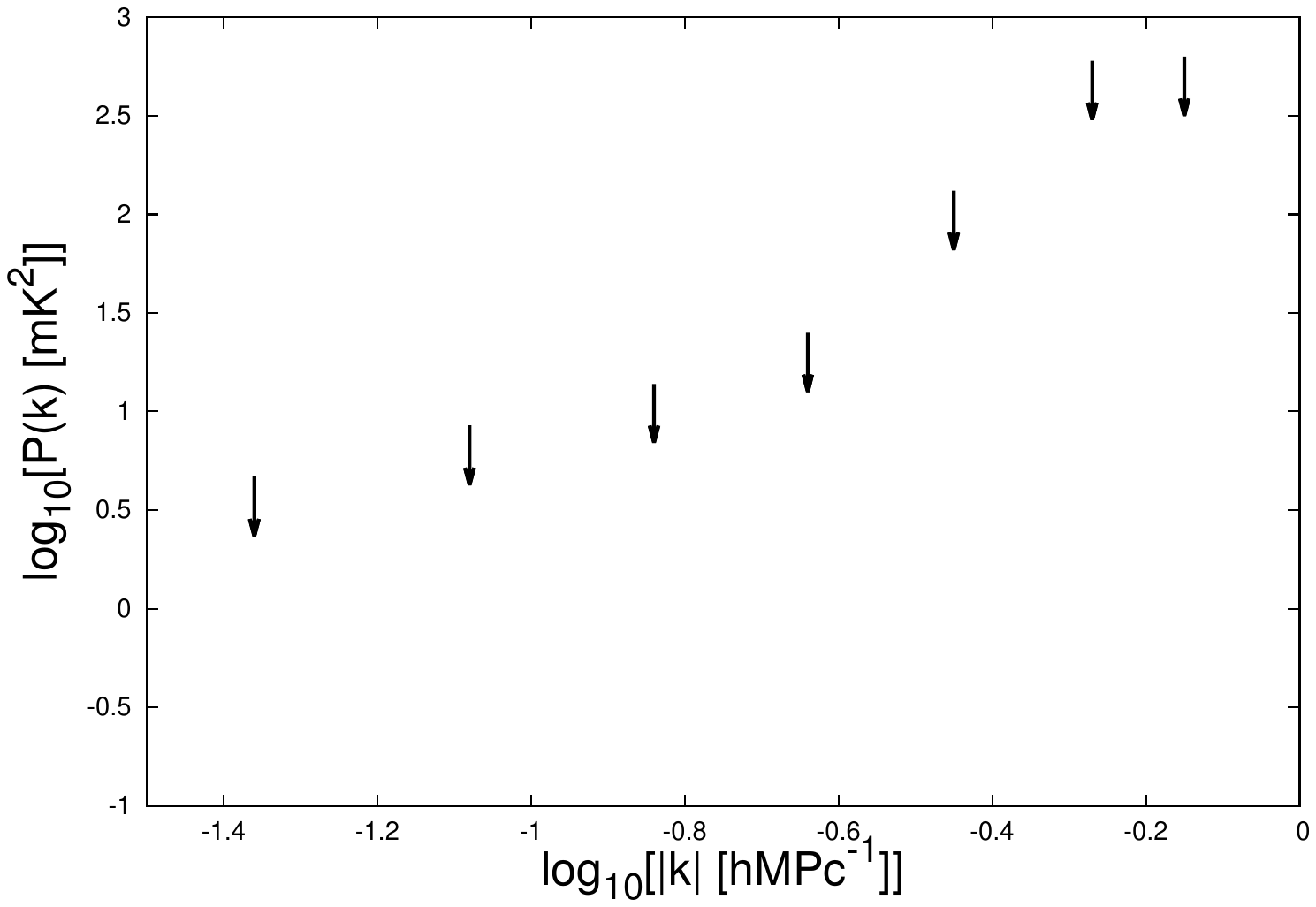} &
\hspace{-1.6cm}
\includegraphics[width=80mm]{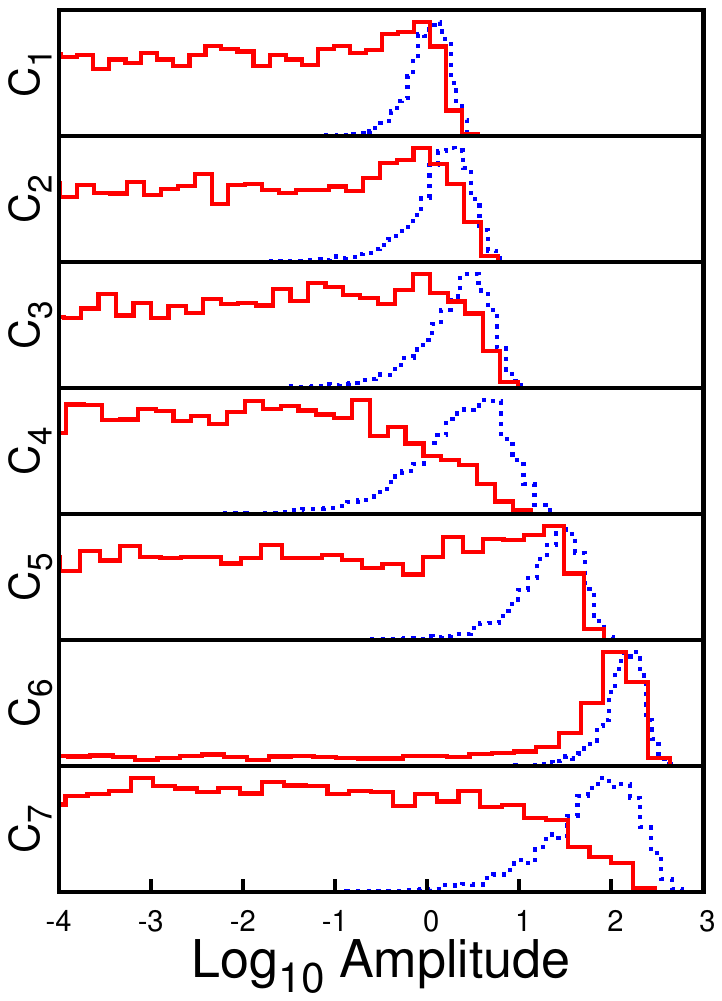} \\
\end{array}$
\end{center}
\caption{(Left) Maximum likelihood reconstructed cylindrical power spectrum for Simulation 1 (a single point source).  There is significant power only in the lowest k-mode, corresponding to the offset term in our model k-cube.  (Center) Recovered values for the spherical power spectrum.  No significant power is detected in any of the coefficients, indicating that all the power in the simulation is correctly modelled by the offset term in the k-cube.  (Right) One dimensional marginalised posteriors for the 7 spherical power spectrum coefficients using priors that are uniform in the amplitude (blue lines) and uniform in the log of the amplitude (red lines).}
\label{figure:Sim1PS}
\end{figure*}

In Figure \ref{figure:Baselines} (red lines) we show the injected real (left) and imaginary (right) signal for the longest (top) and shortest (bottom) baselines for simulation 1, containing a single point source observed by the 37 element array shown in Fig. \ref{figure:SimsSetup} (top).  Noticeably, the baselines show structure as a function of frequency, despite the fact that the injected source has a flat spectrum.  This is simply a result of the baselines sampling a range of UV coordinates, and therefore signal phase, as a function of frequency.  For this simulation we have not included the linear or quadratic terms in our model, opting to use only the offset and the set of the 18 lowest frequency Fourier modes.  The structure recovered from our analysis (blue lines) is plotted only for the offset term from this model for the maximum likelihood 
solution.  In all cases this is completely consistent with the injected data, within the level of the added noise.  

\begin{table}
\centering
\caption{$\log$ Evidence values for Simulation 1} 
\centering 
\begin{tabular}{c c} 
\hline\hline 
Model Coefficients & $\log$ Evidence \\[0.5ex] 
\hline 
0 & 0.0\\
$\rho_1$ & -0.3\\
$\rho_2$ & -0.2\\
$\rho_3$ & -0.5\\
$\rho_4$ & 0.0\\
$\rho_5$ & -0.3\\
$\rho_6$ & 0.8\\
$\rho_7$ & -0.3\\
\hline
\end{tabular}
\label{Table:Sim1Evs} 
\end{table} 

In Table \ref{Table:Sim1Evs}, we list the Evidence values for models that include different sets of power spectrum coefficients, where those power spectrum coefficients not listed for each model have been set to 0. This allows us to address the question of model selection in a Bayesian framework. In particular, we can use the difference in the $\log$ Evidence ${\mathcal{Z}}$ between two competing models, which we will denote $\Delta {\mathcal{Z}} = {\mathcal{Z}}_1 - {\mathcal{Z}}_2$, to obtain the probability that the data supports model 1 over model 2 as,

\begin{equation}
P= \frac{\exp{\Delta {\mathcal{Z}}}}{1 + \exp{\Delta {\mathcal{Z}}}}.
\label{Eq:Rval}
\end{equation}
In the following, we will consider $\Delta {\mathcal{Z}} > 3$ to be significant evidence in favour of including a particular power spectrum coefficient in the model, however for more detail on the use of the Evidence in model selection refer to, e.g. \cite{bayesRef}.  Given this threshold, we can see none of the power spectrum coefficients result in a significant increase in the Evidence, indicating that the included offset term is sufficient to model the entire signal present in the data, simply as a result of defining our model cube in wavelengths, and then projecting this onto our sampled data points.

In Figure \ref{figure:Sim1PS} we show the 1 dimensional marginalised posterior parameter estimates for the 7 spherical power spectrum coefficients when included simultaneously in the model.  All coefficients are consistent with zero, consistent with the change in the Evidence when considering each term individually.  We note here that the most significant increase in the Evidence came from including $\rho_6$, and, from Figure \ref{figure:Sim1PS},  we can see the posterior for the 6th coefficient shows a marginal probability of there being power in the data set at that scale.  This same feature is present at similar significance in simulations 2-3 however, which use the same thermal noise realisation, implying that this is simply a fluctuation in the uncorrelated noise.

\subsection{Results for Simulations 2-3}

\begin{table}
\centering
\caption{$\log$ Evidence values for Simulations 2-3} 
\centering 
\begin{tabular}{c c c} 
\hline\hline 
Model Coefficients & Sim 1 $\log$ Evidence & Sim 2 $\log$ Evidence \\[0.5ex] 
\hline 
0 & 0.0 & 0.0\\
$\rho_2$ & 30.2 & 30.0\\
$\rho_2, \rho_3$ & 35.9 & 36.0\\
$\rho_2, \rho_3, \rho_4$ & 37.1 & 37.3\\
\hline
\end{tabular}
\label{Table:Sim2Evs} 
\end{table} 

As for Simulation 1, Table \ref{Table:Sim2Evs} lists the Evidence for models that include different sets of power spectrum coefficients for simulations 2 and 3. In this case, as we increase the number of coefficients in the model, we only list the particular set that maximises the Evidence.  We find that the Evidence values are consistent between Simulations 2 and 3, and conclude that only 2 spherical power spectrum coefficients have been detected with significance above our threshold. Figure \ref{figure:Sim1PS} shows the results from the analysis of Simulations 2 and 3 using the analytic marginalisation process described in Section \ref{Section:Marginalisation} when including all 7 spherical power spectrum coefficients simultaneously in the model.  In particular, we show the one dimensional marginalised posteriors for the spherical power spectrum coefficients from simulations 2 (middle plot) and 3 (right plot) when using priors that are uniform in the amplitude of the coefficient (blue line) and uniform in the log of the amplitude (red line).  We indicate the 2 coefficients that we consider to be detections in Fig. \ref{figure:Sim1PS} (left) as the points with uncertainties, while the remaining five amplitudes are taken to be 2$\sigma$ upper limits obtained using the prior that is uniform in the amplitude of the coefficient and are represented as arrows in this plot.
All the coefficients for both simulations are consistent with the values obtained from the input cube within $2\sigma$ uncertainties.  Critically, the results from both simulations are completely consistent with one another, indicating that the addition of a significant flat spectrum continuum component, with power 8 orders of magnitude greater than the EoR signal, did not impact our ability to correctly infer the properties of the power spectrum.  

\begin{figure*}
\begin{center}$
\begin{array}{ccc}
\hspace{-0.5cm}
\includegraphics[width=100mm]{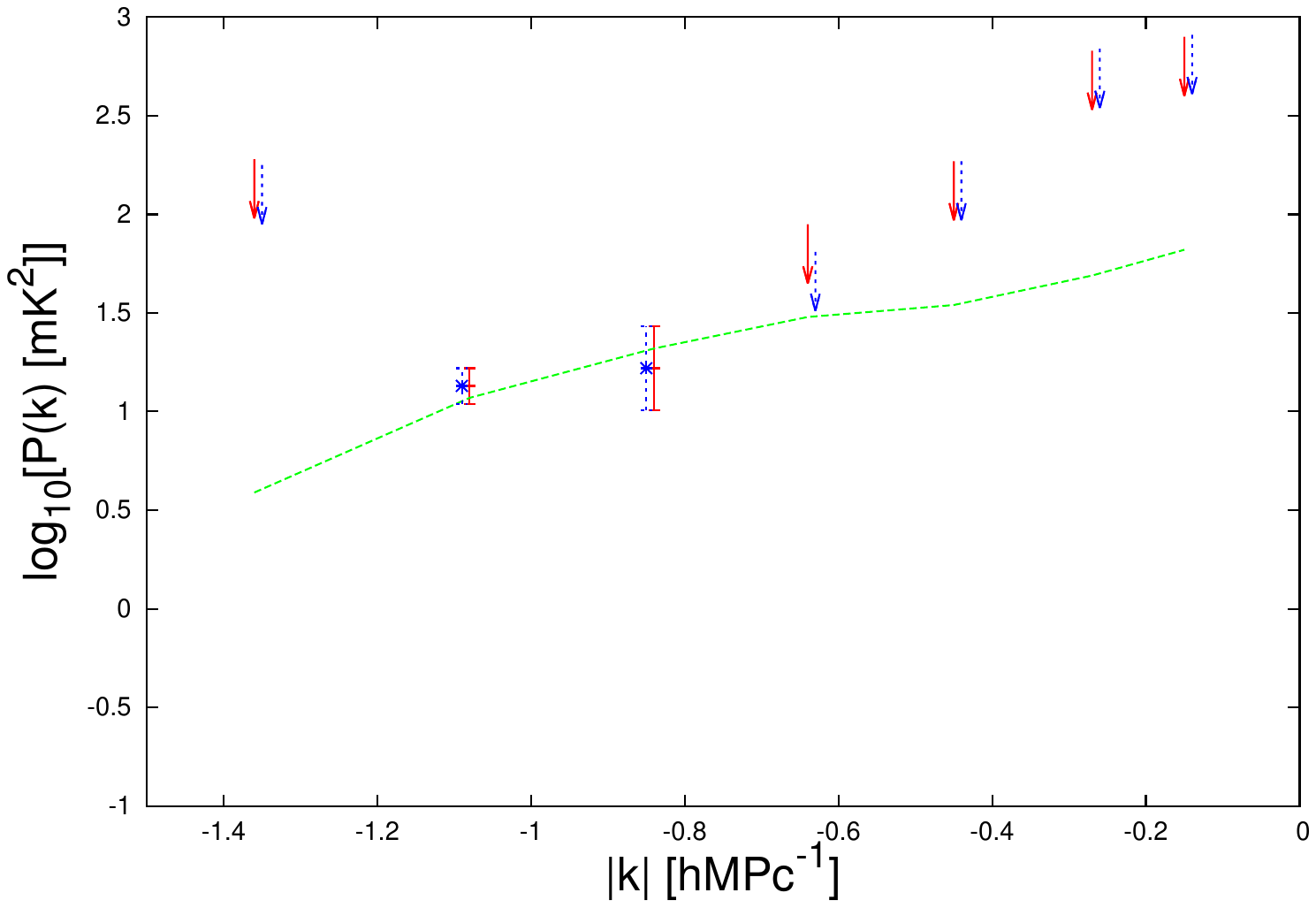} &
\hspace{-1.8cm}
\includegraphics[height=82mm, width=100mm]{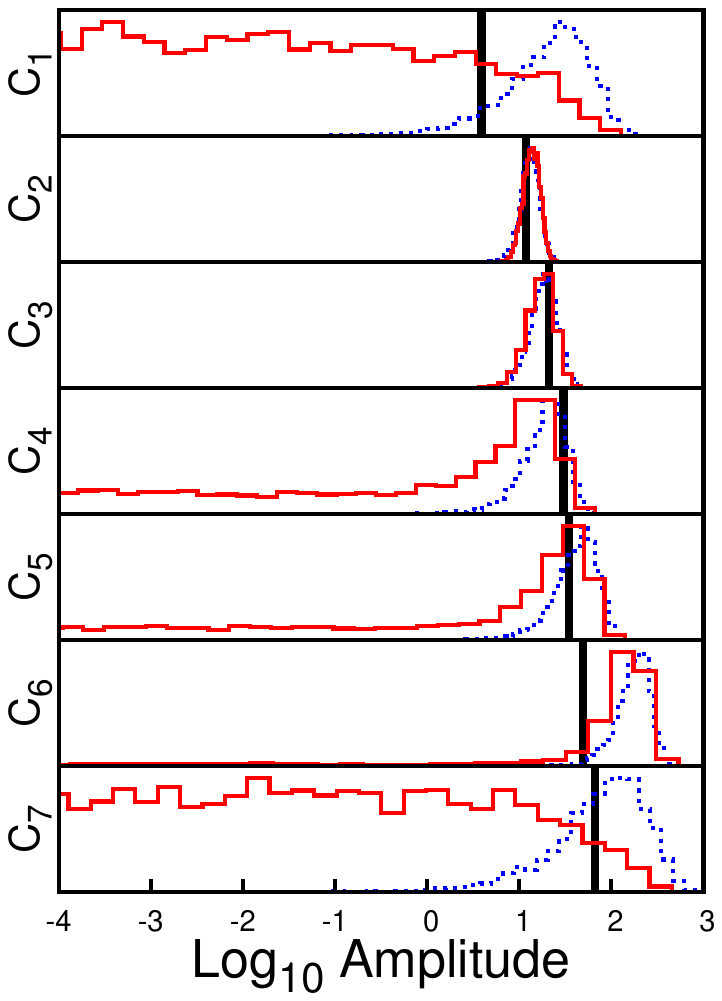} &
\hspace{-6.0cm}
\includegraphics[height=82mm, width=100mm]{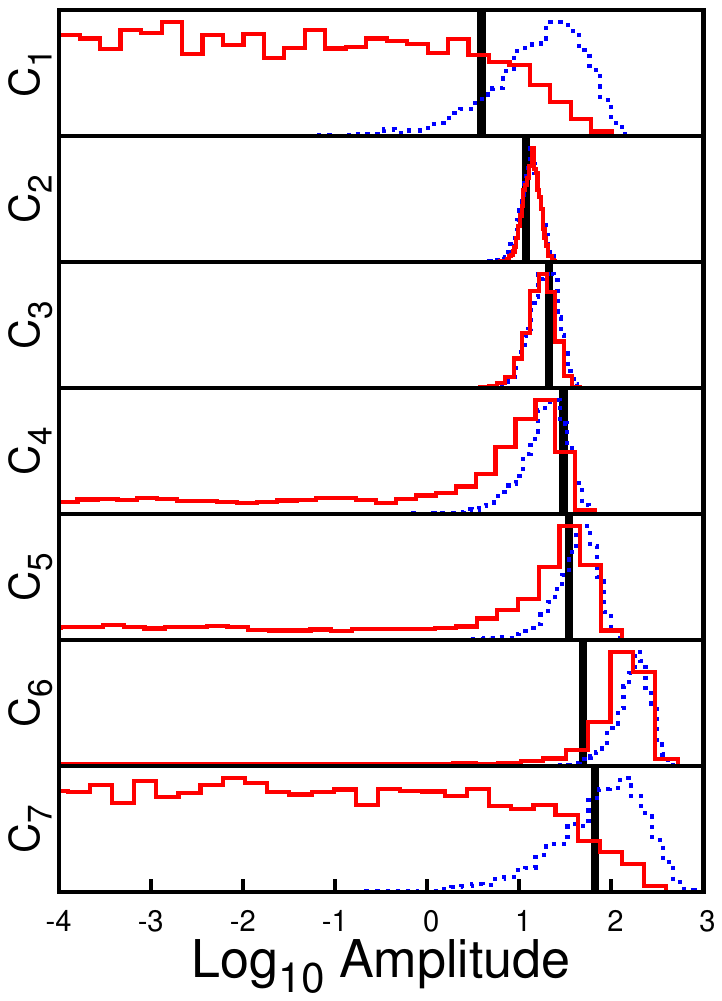} \\
\end{array}$
\end{center}
\caption{(Left) Injected (green line) and recovered values for the spherical power spectrum for Simulation 2 (blue points) and Simulation 3 (red points).  Arrows represent 2$\sigma$ upper limits obtained using a uniform prior on the amplitudes of the coefficients, while points with error bars are the parameter estimates and 1$\sigma$ uncertainties for terms detected using the Log prior.  (Middle) One dimensional marginalised posteriors for the 7 spherical power spectrum coefficients from Simulation 2 using priors that are uniform in the amplitude (blue lines) and uniform in the log of the amplitude (red lines). (Right) As for the middle plot, but for Simulation 3. \label{figure:Sim2PS}}
\end{figure*}

Of note is that, compared to Simulation 1, the upper limits for the lowest spherical power spectrum bin in Fig. \ref{figure:Sim2PS} are considerably worse, despite the fact that the thermal noise realisation is exactly the same between these sets of simulations.  That is because for these two (and subsequent) simulations, we are including the quadratic in our model as a proxy for large spectral scale fluctuations in the data.  The quadratic is most strongly correlated with the largest spatial scales in our EoR signal model, so it decreases our sensitivity to terms in the corresponding lowest k-bin. Including higher order polynomial terms in the fit will extend this effect into the higher k-bins, as cubics, or beyond,  will be more strongly correlated with the higher frequency modes in the model. We stress, however, that this is not a shortcoming of our analysis method.  Fully incorporating the covariance between the low-frequency terms, assumed to be dominated by foregrounds, and the higher-frequency modes of interest is critical in order to obtain unbiased estimates of the EoR power spectrum.

\begin{figure*}
\begin{center}$
\begin{array}{ccc}
\hspace{-0.5cm}
\includegraphics[width=100mm]{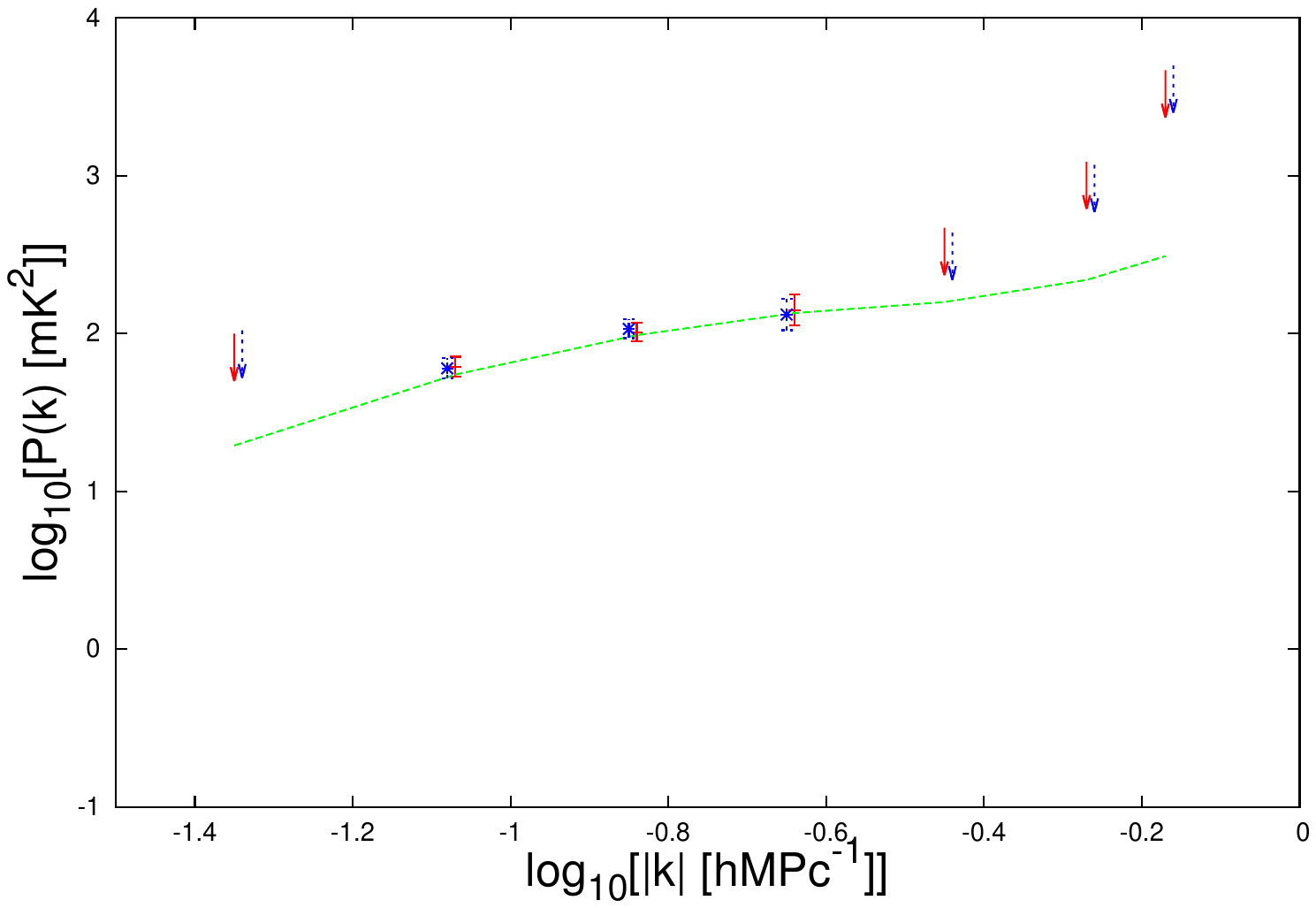} &
\hspace{-1.8cm}
\includegraphics[height=82mm, width=100mm]{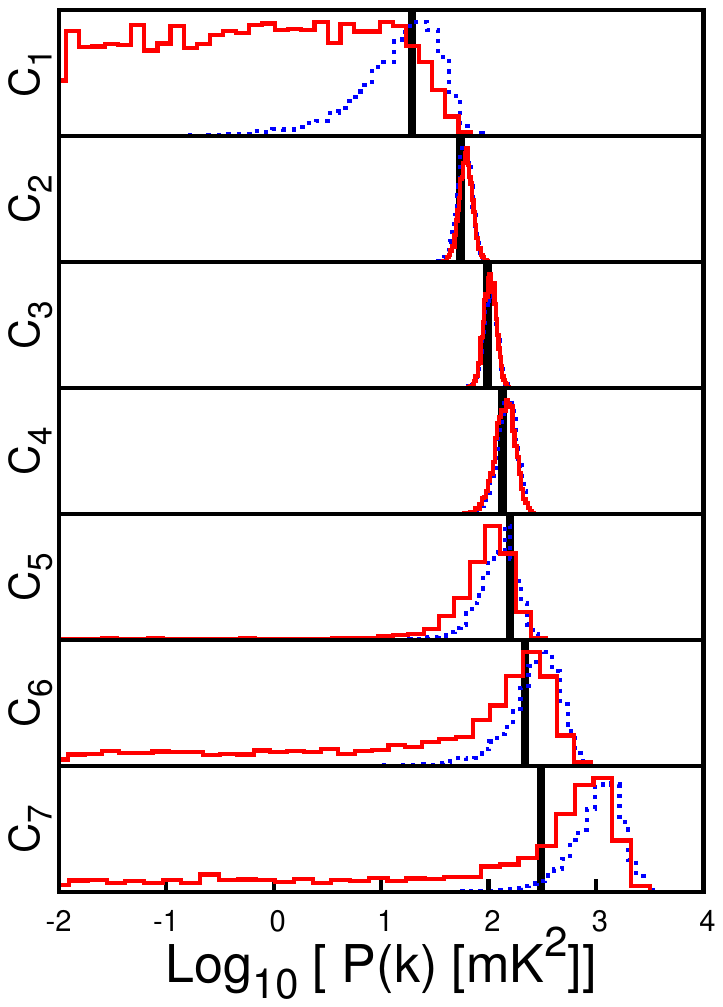} &
\hspace{-6.0cm}
\includegraphics[height=82mm, width=100mm]{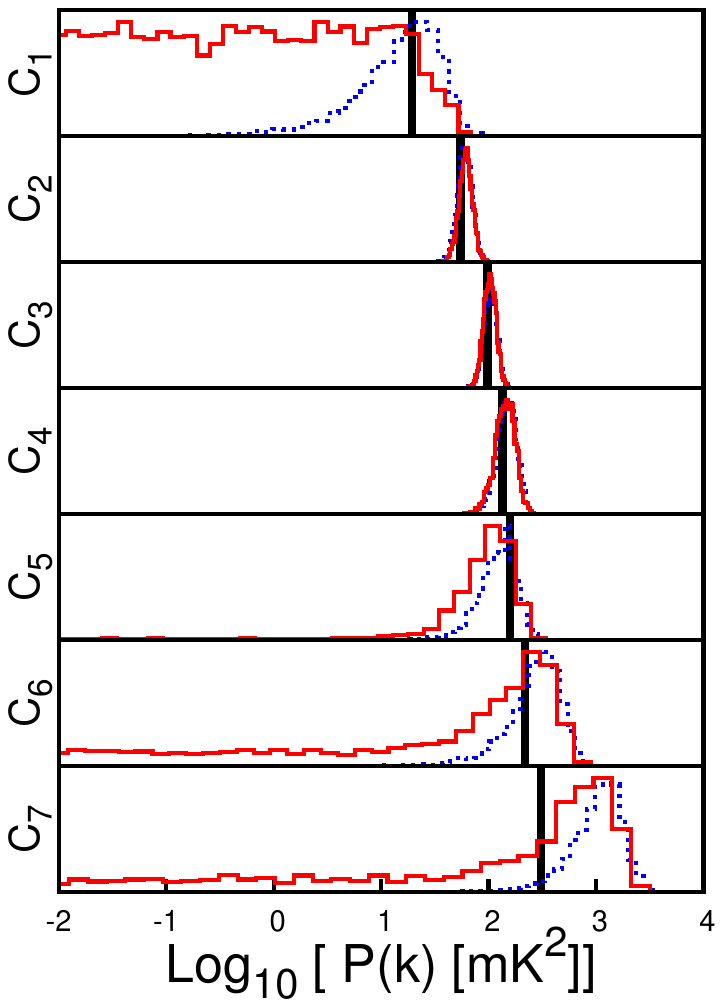} \\
\end{array}$
\end{center}
\caption{(Left) Injected (green line) and recovered values for the spherical power spectrum for Simulation 4 (blue points) and Simulation 5 (red points).  Arrows represent 2$\sigma$ upper limits obtained using a uniform prior on the amplitudes of the coefficients, while points with error bars are the parameter estimates and 1$\sigma$ uncertainties for terms detected using the Log prior.  (Middle) One dimensional marginalised posteriors for the 7 spherical power spectrum coefficients from Simulation 4 using priors that are uniform in the amplitude (blue lines) and uniform in the log of the amplitude (red lines).  (Right) As for the middle plot, but for Simulation 5. \label{figure:Sim4PS}}
\end{figure*}

\subsection{Results for Simulations 4-5}

Figure \ref{figure:Sim4PS} shows the one dimensional marginalised posteriors for the spherical power spectrum coefficients from simulations 4 (middle plot) and 5 (right plot) when using priors that are uniform in the amplitude of the coefficient (blue line) and uniform in the log of the amplitude (red line).  As we are now using the GHS to perform the sampling, we no longer obtain the evidence for different sets of coefficients.  As such, we consider a power spectrum coefficient `detected' when the posterior is not consistent with $\log_{10}$ amplitudes less than -2 when using a logarithmic prior, which, given the noise level in the simulation, is equivalent to being consistent with zero.
Compared to the 37 element array, the 61 element array provides much greater constraints on the 2nd and 3rd spherical power spectrum coefficients and provides a detection of the 4th coefficient.  The remaining terms, however, are still consistent with zero when using logarithmic priors, and so we consider these only 2$\sigma$ upper limits, obtained using the uniform priors.  All the coefficients for both simulations are consistent with the values obtained from the input cube, within $2\sigma$ uncertainties. 

\section{Comparison with Other Power Spectrum Estimators}
\label{Section:Comparison}

While a complete summary of the literature on 21-cm power spectrum estimation is outside the scope of this work, it is useful to describe the general classes of estimators and put our work in context.  We first compare with the (more developed) non-Bayesian estimators in the literature in Section \ref{Section:Comparison-NoBayes}, and then with two recently proposed Bayesian estimators in Section \ref{Section:Comparison-Bayes}.

\subsection{Non-Bayesian Approaches}
\label{Section:Comparison-NoBayes}

Morales et al. (in prep.) propose a useful classification of power spectrum estimators, dividing the literature into ``measured'' and ``reconstructed'' sky approaches.  Measured power spectrum estimators effectively return the power spectrum of the sky with no attempt to remove spectral features introduced by the chromatic response of the interferometer.  These estimators are typified by the PAPER-style ``delay spectrum'' approach \citep{2012ApJ...756..165P}, which never coherently combines measurements from different baselines.  Reconstructed sky estimators, on the other hand, coherently combine measurements from all baselines (generally by ``gridding'' visibilities into a UV plane) and return the best estimate of the true sky, free from the effects of the interferometer.  The analyses used by MWA \citep{2016ApJ...825..114J} and LOFAR \citep{2017ApJ...838...65P} are archetypes of this approach.  Interestingly, both these analysis types show the ``wedge'' feature of baseline-length-dependent spectral contamination of smooth spectrum foregrounds, but with different properties and sensitivities to calibration errors.  See Morales et al., (in prep.), for more details. 

In terms of this classification, our estimator is clearly a \emph{reconstructed} sky estimator.  Our model for the data (given in its final form in Equation \ref{Eq:FinalModel}) is simultaneously compared with all of the data in UV space, meaning measurements from all baselines are combined.  By doing so, we can effectively remove the wedge feature introduced by the instrument into the measurements.  The work presented here shows that we can remove this feature to within the noise level, as long as the instrument model is perfect; in future work, we will explore the effects of instrument model errors and calibration errors on this technique.

Another useful distinction is between estimators that aim to recover the power spectrum of all emission on the sky and those which specifically aim to recover that from cosmological 21-cm emission. Estimators of the first class often use a prior foreground removal step, independent of power spectrum estimation; the power spectrum estimate therefore contains contributions from both residual foreground emission and cosmological 21-cm emission.  Examples of this class of estimator are the $\epsilon$ppsilon algorithm used in MWA analysis \citep{2016ApJ...825..114J} and the LOFAR power spectrum analysis \citep{2017ApJ...838...65P}.  Alternatively, one can introduce statistical models of the foregrounds or other contaminants (typically through a covariance matrix) and estimate signals that are statistically distinct, in an attempt to isolate 21-cm emission.  Examples of such an approach include the CHiPS pipeline \citep{2016ApJ...818..139T} and the empirical covariance estimation analysis of \cite{2015PhRvD..91l3011D}.  Our analysis also falls into this latter category by jointly estimating the power spectra of foregrounds and 21-cm emission, allowing for isolation of the 21-cm signal even without an explicit foreground removal step.  However, the Bayesian framework represents a major step beyond these existing analyses, in that the full posterior probability distribution for the EoR power spectrum is explored to provide robust uncertainties on the signal estimate. Examples of how our analysis works with more realistic foregrounds are found in \cite{2016MNRAS.462.3069S} and Sims et al., (a, b in review).

\subsection{Bayesian Approaches}
\label{Section:Comparison-Bayes}

Statistically robust recovery of both the power spectral estimates and their uncertainties is essential to avoid spurious or mischaracterised detection of the redshifted 21-cm signal and for enabling reliable inferral of astrophysical constraints from the derived results. A Bayesian approach to estimating the power spectrum of the EoR provides a natural framework within which known uncertainties in the analysis chain can be covariantly propagated through to the power spectral estimates. It thus provides a route through which statically robust results can be achieved. However, incorporating Bayesian statistical elements in the analysis, alone, does not guarantee that this will be the case and, as in any framework, the robustness of the derived results will be sensitive to the particulars of the method employed. Approaches to EoR power spectrum estimation in the literature that incorporate Bayesian statistical elements in their analysis chain (e.g.  \citealt{2015MNRAS.452.1587G, 2016ApJS..222....3Z}) go part-way towards this goal:
\begin{enumerate}                                                                                                                                                                 \item \citet{2015MNRAS.452.1587G} employ a two-stage methodology. First, a generalized morphological component analysis (GMCA) is applied to a Foreground + EoR dataset, producing a residuals dataset comprised of the remaining EoR signal, any unsubtracted foreground signal and noise. The maximum a posteriori (MAP) image cube, using zero-order, gradient or curvature regularisation of the signal coefficients, is calculated from the residuals dataset. The EoR power spectrum is derived from the MAP image cube. For the choice of number of GMCA components, foreground and EoR simulations described in \citet{2015MNRAS.452.1587G} this provides positive results with unbiased estimates of the EoR power spectrum recovered on a range of spatial scales. Nevertheless, the drawback of this approach in realistic applications, where tuning the required degrees of freedom of the foreground model is more difficult, and inherent to all approaches comprised of independent foreground subtraction and power spectral estimation of the power steps, is the potential for signal loss (foreground contamination), if an overly complex (simplistic) foreground model\footnote{This will occur when using either a larger or smaller number of GMCA components than that required to model the foregrounds in a given dataset. In practice, the EoR signal, and thus its correlation with the foreground model for the dataset with a given number of GMCA components, is unknown, further complicating this choice.} is used and the derivation of incorrect uncertainties on the power spectral estimates, if the foreground model is correlated with the EoR signal in the data.                                                                                                                                                                \item In contrast, in \citet{2016ApJS..222....3Z} this drawback is overcome by jointly estimating a model for the EoR signal and the foregrounds, with their foreground model, which is derived via independent component analysis. Joint estimation of the EoR and foreground models allows correlation between the two to be accounted for when estimating the power spectrum and to be reflected in the derived uncertanties of power spectral coefficients on the spatial scales represented in the foreground model. However, recovering the EoR power spectrum is made difficult by i) the relative brightness of the intrinsic foreground signal in comparison to the EoR signal and ii) the mode-mixing effect of the interferometer, which corrupts the intrinsic smoothness of the foreground spectrum, correlating it with the EoR signal in the observed data. As such, a method for estimating the power spectrum of the EoR from interferometric visibility data which is sampled at frequency-dependent $uv$-coordinates and is a function of the frequency-dependent point spread function of the telescope is key to the real-world application of the methodology. In \citet{2016ApJS..222....3Z}, the mode mixing effect of the interferometer is not accounted for, thus further development would be necessary for it to be made applicable to a realistic dataset.
\item In addition, a further difficulty with both approaches presented in \citet{2015MNRAS.452.1587G} and in \citet{2016ApJS..222....3Z} is that both require knowing the data covariance matrix with high precision. This reliance results in a high sensitivity of the recovered power spectrum to inaccuracies in estimates of the effective noise level in the data. Any misestimation of the noise (due either to unmodelled intrinsic small spatial scale power in the signal, or imperfect knowledge of the effective instrumental noise) will translate directly into bias in the recovered power spectral estimates.
\end{enumerate}

In this paper, we develop a new approach that aims to address the respective shortcomings in the aforementioned approaches. As in \citet{2016ApJS..222....3Z}, we jointly estimate models for the EoR signal and foregrounds, allowing us to account for correlation between the two in our derived power spectral estimates. However, in addition, we incorporate instrumental forward modelling in our data model, allowing us to account for the mode mixing effect of the interferometer. We have demonstrated that in the zero-uncertainty limit on the instrumental model, this allows us to estimate the intrinsic power spectrum of the EoR free from instrumental effects.
Further, in our model for the covariance of the data, we also fit for an additional noise term. The primary purpose of this additional noise term is to account for structure in the signal on spatial scales smaller than those Nyquist sampled in the dataset under analysis, and thus not recoverable with perfect fidelity, preventing structure on these scales from leaking into and biasing power spectral estimates on the scales of interest. However, an additional benefit of this parametrisation of our noise model is that, unlike in the approaches discussed above, which assume perfect knowledge of the data covariance matrix and are highly sensitive to inaccuracies in their noise estimates, with any mistakes translating directly to bias in recovered estimates, with this parametrisation, any underestimation of the instrumental noise will be absorbed by the intrinsic noise parameter, preventing bias in the recovered power spectral estimates.

\section{Conclusions}
\label{Section:Conclusions}

We have presented a new Bayesian method for analysing interferometric data in order to estimate the three-dimensional power spectrum of density fluctuations in the neutral hydrogen at the Epoch of Reionization.  

We have described two applications of this method: i) sampling directly from the power spectrum coefficients of the EoR signal by marginalising analytically over the signal coefficients, resulting in a compact parameter space ($\sim$ 10 dimensions) that requires large dense matrix inversions, and ii) sampling from the joint probability density of the power spectrum coefficients and the EoR signal realisation, resulting in large dimensionality ($\sim$ 20000 dimensions), but eliminating all matrix-matrix multiplications and costly matrix inversions from the likelihood calculation entirely, replacing them with matrix-vector operations and diagonal matrix inversions.  In this case, we performed the sampling process using a Guided Hamiltonian Sampler (B18) which provides an efficient means of sampling in large numbers of dimensions (potentially $> 10^6$).

We then used a series of simulations to show that both approaches presented allow for a reconstruction of the EoR power spectrum that is consistent with the model injected into the simulation in both high and low signal to noise regimes.  When adding a simple, flat spectrum continuum model, the power in which was $\sim 10^8$ times greater than the EoR signal, we showed that the estimates of the power spectrum were unaffected, despite no prior knowledge of the value or distribution of source amplitudes continuum in the continuum sky being used in the analysis.

In \cite{2016MNRAS.462.3069S} this approach has been used to estimate the three-dimensional power spectrum of interferometric data sets in the presence of astrophysically realistic foregrounds. Here, it was found that these foregrounds contain power on all scales of interest, and that simultaneous estimation of both the EoR and foregrounds is important in order to obtain statistically robust estimates of the EoR power spectrum.  Biased results, and thus biased astrophysical parameter estimates, will be obtained from methodologies that do not incorporate this covariance.  Thus, methods such as those discussed in this work will be essential as we move towards the eventual detection of the EoR and attempt to infer astrophysical conclusions about galaxy formation in the early Universe.

\section{Acknowledgements}

This work was performed using the Darwin Supercomputer of the University of Cambridge High Performance Computing Service (http://www.hpc.cam.ac.uk/), provided by Dell Inc. using Strategic Research Infrastructure Funding from the Higher Education Funding Council for England and funding from the Science and Technology Facilities Council.
PMS is supported by the INFN IS PD51 ``Indark".
PHS and JCP acknowledge support from NSF award \#1636646.

\end{document}